\documentclass[preprint,12pt]{elsarticle}

\usepackage{amssymb}

\usepackage{amsmath}

\usepackage[colorlinks=true, linkcolor=blue, citecolor=red, urlcolor=magenta]{hyperref}
\usepackage{booktabs}
\usepackage{makecell} 
\usepackage{graphicx}
\usepackage{multirow}

\begin{document}
	
	\begin{frontmatter}
		
\title{Interpolation-supplemented lattice Boltzmann simulation of thermal convection on non-uniform meshes
\footnote{\href{https://doi.org/10.1016/j.ijheatmasstransfer.2025.127790}{DOI: 10.1016/j.ijheatmasstransfer.2025.127790}}
\footnote{\copyright \ 2025. This manuscript version is made available under the CC-BY-NC-ND 4.0 license http://creativecommons.org/licenses/by-nc-nd/4.0/}
}
		\author[inst1,inst2]{Ao Xu\corref{cor1}}
		\cortext[cor1]{\*Corresponding author: \href{mailto:axu@nwpu.edu.cn}{axu@nwpu.edu.cn} (Ao Xu)}
		\author[inst1]{Zheng Zhao}
		\author[inst1]{Ben-Rui Xu}
		\author[inst1]{Li-Sheng Jiang}
		\affiliation[inst1]{organization={Institute of Extreme Mechanics, School of Aeronautics},
			addressline={Northwestern Polytechnical University}, 
			city={Xi’an},
			postcode={710072}, 
			country={PR China}}
		
		\affiliation[inst2]{organization={National Key Laboratory of Aircraft Configuration Design},
			addressline={Key Laboratory for Extreme Mechanics of Aircraft of Ministry of Industry and Information Technology}, 
			city={Xi’an},
			postcode={710072}, 
			country={PR China}}
\begin{abstract}
We present a systematic evaluation of an interpolation-supplemented lattice Boltzmann method (ISLBM) for simulating buoyancy-driven thermal convection on non-uniform meshes. 
The ISLBM extends the standard lattice Boltzmann framework by incorporating quadratic interpolation during the streaming step, enabling flexible mesh refinement near solid boundaries while maintaining algorithmic simplicity and parallel scalability. 
The method is implemented for a two-dimensional side-heated cavity at high Rayleigh numbers $10^6\leq Ra \leq 10^8$, and for a three-dimensional side-heated cavity at $10^5\leq Ra \leq 10^7$, with the Prandtl number fixed at $Pr=0.71$. 
Benchmark results show that the ISLBM accurately captures thermal and velocity boundary layers, yielding Nusselt and Reynolds numbers in close agreement with high-fidelity reference data. 
Grid-convergence studies demonstrate nearly third-order accuracy for global quantities and about second-order for local fields.
We further assess the computational performance of the in-house LBM solver against two open-source solvers: Nek5000 based on the spectral element method, and OpenFOAM based on the finite volume method. 
Performance metrics, including million lattice updates per second (MLUPS) and wall-clock time per dimensionless time unit (WCTpDT), indicate that the ISLBM offers one to three orders of magnitude higher efficiency in large-scale simulations. 
On GPU architectures, the ISLBM retains high computational performance: throughput on non-uniform meshes reaches 60–70\% of that on uniform meshes in terms of MLUPS, while the cost in WCTpDT is about three times higher.
These results highlight the potential of interpolation-based LBM approaches for high-fidelity simulations of thermal convection on non-uniform meshes, providing a robust foundation for future extensions to turbulent flows.
\end{abstract}
		
		
		
		
	\end{frontmatter}
	
		
		
		\section{Introduction} \label{sec:introduction}

The lattice Boltzmann method (LBM) is a mesoscopic numerical approach to computational fluid dynamics \cite{Chen1998,Aidun2010}, rooted in the Boltzmann kinetic theory \cite{xu2015direct,xu2015directbook}. 
Instead of directly solving the macroscopic Navier-Stokes equations, the LBM models the evolution of particle distribution functions on a discrete lattice \cite{qian1992lattice,chen1992recovery}. 
From these evolution equations, macroscopic flow variables such as velocity and pressure are recovered.
This framework provides access to continuum-scale hydrodynamics while retaining mesoscopic fidelity.
Early developments of the LBM were inspired by lattice gas automata, but a rigorous connection to the Boltzmann equation was later established by He and Luo \cite{he1997theory}, who derived the lattice Boltzmann equation from the continuous Bhatnagar-Gross-Krook formulation. 
Over the past three decades, the LBM has matured into an efficient tool for simulating a wide range of complex flow phenomena. 
Its inherent algorithmic simplicity, local collision-streaming structure, and natural suitability for parallel computation \cite{xu2017accelerated,xu2023multi,xu2024particle} make it particularly effective for simulating fluid flows and associated transport processes. 
As a result, the LBM has achieved broad success across diverse applications, including turbulent and multiphase flow simulations \cite{chen2014a,li2016lattice,Calzavarini2019,Patocka2020,Patocka2022,Wu2024,Zhang2024,Liu2025a,Liu2025b}. 
This success has positioned LBM as a compelling alternative to conventional Navier-Stokes-based solvers, particularly for incompressible flows in the continuum regime.

Our particular interest lies in using the LBM to simulate thermal convection, specifically the coupled fluid flow and heat transfer processes characteristic of buoyancy-driven systems \cite{Xia2013,Xia2023,xia2025some}. 
This fundamental mechanism underpins a wide range of natural and engineered systems, including atmospheric and oceanic convection \cite{xia2025some}, thermal management in fuel cells and flow batteries \cite{xu2017lattice}, thermal protection in nuclear reactors \cite{ni2025Magnetohydrodynamics}, and thermal convection in mechanical devices subjected to vibrations \cite{Wang2020}. 
Accurately simulating thermal convection at high Rayleigh numbers ($Ra$, defined later) remains a major computational challenge, as strong buoyancy forces significantly influence the flow and heat transfer dynamics. 
The difficulty arises from the formation of thin thermal boundary layers, which demand much finer resolution near solid boundaries than in the bulk flow. 
As a result, uniform meshes quickly become computationally prohibitive, since they enforce fine resolution across the entire domain. 
To overcome this limitation, it is necessary to extend the conventional LBM (typically formulated on uniform meshes) to non-uniform meshes refined near solid boundaries, where steep velocity and temperature gradients occur. 
Such spatially adaptive meshing enables adequate resolution of near-wall structures, such as thermal boundary layers, while preserving computational efficiency in the bulk flow region.

Several extensions of the standard LBM from uniform meshes to non-uniform meshes have been developed to resolve near-wall gradients without incurring excessive computational cost. 
A prominent example is the interpolation-supplemented LBM (ISLBM) \cite{he1996some,he1997some,he1997error}, which incorporates interpolation schemes during the streaming step to enable off-lattice population transfers, thereby allowing non-uniform lattice spacing while retaining the simplicity of the standard formulation.
Another widely studied approach is the finite-volume LBM \cite{Nannelli1992,Peng1999,Xi1999,Ubertini2003}, which reformulates the lattice Boltzmann equation within a control-volume framework, permitting flexible mesh structures while preserving its kinetic foundation. 
In addition, mesh refinement strategies, such as block-structured multi-domain methods \cite{Filippova1998,Yu2002} and multilevel (multigrid) methods \cite{Tolke2002,Mavriplis2006}, have been extensively employed to resolve localized flow features at varying spatial resolutions. 
These methodological advances are particularly relevant as the LBM is increasingly applied to simulate thermal convection at high Rayleigh numbers, where both large-scale circulations and thin boundary layers must be resolved.
Despite their potential, the performance of these methods in terms of numerical accuracy, stability, and computational efficiency for thermally driven flows at high Rayleigh numbers have not been systematically assessed. 
Therefore, a comprehensive evaluation of their accuracy, numerical stability, and efficiency is essential for advancing high-fidelity LBM simulations of buoyancy-driven flows.

Due to its conceptual simplicity and its direct extension of the standard streaming step, we focus here on evaluating the ISLBM. 
Specifically, we implement and assess the ISLBM for simulating buoyancy-driven thermal convection at moderately high Rayleigh numbers on non-uniform meshes. 
Through systematic benchmarks, we aim to quantify its accuracy and efficiency, thereby laying the groundwork for its future application to turbulent thermal systems. 
The rest of this paper is organized as follows. 
In Section \ref{sec:methods}, we present the numerical details of the ISLBM for simulating thermal convection on non-uniform meshes. 
In Sections \ref{sec:SHC-2D} and \ref{sec:SHC-3D}, we present results on laminar thermal convection in a two-dimensional (2-D) and a three-dimensional (3-D) side-heated cavity, respectively. 
In Section \ref{sec:conclusion}, the main findings of the present work are summarized.

\section{Numerical methods} \label{sec:methods}
				
\subsection{The standard LBM for simulating thermal convection on uniform meshes}
		
We perform direct numerical simulations of thermal convection under the Boussinesq approximation. 
The fluid flow is assumed to be incompressible, and temperature is treated as an active scalar that influences the velocity field through buoyancy. 
Viscous dissipation and compression work are neglected, and all transport coefficients are assumed constant. 
The governing equations for the coupled fluid flow and heat transfer are

		\begin{equation}
			\nabla \cdot \mathbf{u}=0  
		\end{equation}
		\begin{equation}
			\frac{\partial \mathbf{u}}{\partial t}+\mathbf{u} \cdot \nabla \mathbf{u}=-\frac{1}{\rho_0} \nabla P+\nu \nabla^2 \mathbf{u}+g \beta\left(T-T_0\right) \hat{\mathbf{y}} 
		\end{equation}
		\begin{equation}
			\frac{\partial T}{\partial t}+\mathbf{u} \cdot \nabla T=\alpha \nabla^2 T 
		\end{equation}
where $\mathbf{u}$, $P$ and $T$ denote the velocity, pressure, and temperature fields, respectively. 
The reference density and temperature are denoted by $\rho_0$ and $T_0$.
$\nu$, $\beta$ and $\alpha$ are the kinematic viscosity, thermal expansion coefficient, and thermal diffusivity of the fluid, respectively. 
The unit vector $\hat{\mathbf{y}}$ points in the direction of gravity. 
To non-dimensionalize the equations, we apply the following scalings:
		\begin{equation}
			\begin{aligned}
				&\mathbf{x}^{*}=\mathbf{x}/H,\quad t^{*}=t/\sqrt{H/\left(\beta g\Delta_{T}\right)},\quad \mathbf{u}^{*}=\mathbf{u}/\sqrt{\beta gH\Delta_{T}}\\
				&P^{*}=P/\left(\rho_{0} g\beta\Delta_{T} H\right),\quad T^{*}=\left(T-T_{0}\right)/\Delta_{T}
				\label{eq:2}
			\end{aligned}
		\end{equation}
where $\Delta_T$ is the temperature difference between the hot and cold walls. 
The characteristic length scale is the cavity height $H$, the characteristic time scale is the free-fall time $t_f = \sqrt{H/(\beta g \Delta_T)}$, and the characteristic velocity scale is the free-fall velocity $u_f = \sqrt{H \beta g \Delta_T}$. 
Unless otherwise stated, dimensionless variables are denoted with a superscript star.

Using these scalings, the dimensionless governing equations become
		\begin{equation}
			\nabla\cdot\mathbf{u}^* = 0  
		\end{equation}
		\begin{equation}
			\frac{\partial\mathbf{u}^*}{\partial t^*} + \mathbf{u}^*\cdot\nabla\mathbf{u}^* = -\nabla P^* + \sqrt{\frac{Pr}{Ra}}\nabla^2\mathbf{u}^* + T^*\mathbf{\hat{y}} 
		\end{equation}
		\begin{equation}
			\frac{\partial T^*}{\partial t^*} + \mathbf{u}^*\cdot\nabla T^* = \sqrt{\frac{1}{PrRa}}\nabla^2T^*
		\end{equation}
Two dimensionless parameters arise in the system: the Rayleigh number ($Ra$) and the Prandtl number ($Pr$), defined as
		\begin{equation}
			Ra=\frac{g\beta\Delta_TH^3}{\nu \alpha},\quad Pr=\frac{\nu}{\alpha} 
		\end{equation}
The Rayleigh number quantifies the driving buoyancy force relative to viscous and thermal diffusion, while the Prandtl number represents the ratio of momentum diffusivity to thermal diffusivity.

To solve the thermal convection problem described above, we adopt a double-distribution-function (DDF) LB model \cite{xu2017accelerated,xu2019lattice,wang2013lattice,contrino2014lattice,chai2013lattice,chai2020multiple}. 
Specifically, a D2Q9 lattice in two dimensions or a D3Q19 lattice in three dimensions is employed to solve the Navier–Stokes equations for fluid flow, while a D2Q5 lattice in two dimensions or a D3Q7 lattice in three dimensions is used to solve the energy equation for heat transfer. 		
To enhance numerical stability, a multi-relaxation-time (MRT) collision operator is employed for both the density and temperature distribution functions. 
The evolution equation for the density distribution function is 
		\begin{align}
			f_i(\mathbf{x}+\mathbf{e}_i\delta_t,t+\delta_t) - f_i(\mathbf{x},t) &= -\left(\mathbf{M}^{-1}\mathbf{S}\right)_{ij}\left[\mathbf{m}_j(\mathbf{x},t) - \mathbf{m}_j^{(\text{eq})}(\mathbf{x},t)\right] + \delta_tF_i^{\prime} \notag \\
			&\quad (i=0,1,\cdots,q-1)
			\label{eq:f}
		\end{align}
where $f_{i}$ is the density distribution function, $\mathbf{x}$ is the spatial position, $t$ is the time, and $\delta_{t}$ is the time step. 
$\mathbf{e}_{i}$ is the discrete velocity vector in the $i$th direction. 
For the D2Q9 lattice, $q=9$; for the D3Q19 lattice, $q=19$. 
The equilibrium moments $\mathbf{m}^{(\text{eq})}$ are given by, for the D2Q9 lattice
		\begin{equation}
				\mathbf{m}^{(\text{eq})}_{\text{D2Q9}} = 
				\rho \bigl[ 1,\ -2 + 3(u^2 + v^2), \ 1 - 3(u^2 + v^2), \ u, \ -u, \ v, \ -v, \ u^2 - v^2, \ uv \bigr]^T
				\label{eq:equilibriumMomentsD2Q9}
		\end{equation}
and for the D3Q19 lattice,
\begin{equation}
    \begin{split}
\mathbf{m}^{(\text{eq})}_{\text{D3Q19}}
=\rho \bigg[
& 1,\ -11+19|\mathbf{u}|^{2},\ 3-\frac{11}{2}|\mathbf{u}|^{2},\ u,\ -\frac{2}{3}u,\ v,\ -\frac{2}{3}v, \ w,\\
& -\frac{2}{3}w, \ 2u^{2}-v^{2}-w^{2},\ -\frac{1}{2}(2u^{2}-v^{2}-w^{2}), \ v^{2}-w^{2}, \\
& -\frac{1}{2}(v^{2}-w^{2}), \ uv, \ vw, \ uw, \ 0,\ 0, \ 0 \bigg]^{T} \\
    \end{split}
\end{equation}
The forcing term $F_i^{\prime}$ on the right-hand side of Eq. (\ref{eq:f}) is computed as $\mathbf{F}^{\prime}=\mathbf{M}^{-1}\left(\mathbf{I}-\mathbf{S}/2\right)\mathbf{M}\mathbf{\tilde{F}}$, where $\mathbf{M}\mathbf{\tilde{F}}$ is given by Guo et al. \cite{guo2002discrete,guo2008analysis}.
For the D2Q9 lattice,
		\begin{equation}
			\mathbf{M}\tilde{\mathbf{F}}_{D2Q9}=\begin{bmatrix}0,6\mathbf{u}\cdot\mathbf{F},-6\mathbf{u}\cdot\mathbf{F},F_x,-F_x,F_y,-F_y,2uF_x-2vF_y,uF_x+vF_y\end{bmatrix}^T 
		\end{equation}
and for the D3Q19 lattice,
\begin{small}
\begin{equation}
    \begin{split}
\mathbf{M}\tilde{\mathbf{F}}_{D3Q19}
& =\bigg[ 0,\   38\mathbf{u}\cdot\mathbf{F},\  -11\mathbf{u}\cdot\mathbf{F},\  F_{x},\  -\frac{2}{3}F_{x},\  F_{y},\  -\frac{2}{3}F_{y},  \ F_{z}, \ -\frac{2}{3}F_{z}, \\
& \ 4uF_{x}-2vF_{y}-2wF_{z},\  -2uF_{x}+vF_{y}+wF_{z},  \ 2vF_{y}-2wF_{z}, \\
& \ -vF_{y}+wF_{z},  \ uF_{y}+vF_{x},  \ vF_{z}+wF_{y},  \ uF_{z}+wF_{x},  \ 0, \ 0, \ 0  \bigg]^{T}
    \end{split}
\end{equation}
\end{small}
Here, the body force is defined as $\mathbf{F} = \rho g \beta (T - T_0) \hat{\mathbf{y}}$. 
The macroscopic density $\rho$ and velocity $\mathbf{u}$ are obtained via $\rho=\sum_{i=0}^{q-1}f_i$, $\mathbf{u}=\frac{1}{\rho}(\sum_{i=0}^{q-1}\mathbf{e}_i f_i+\mathbf{F}/2)$.

The evolution equation for the temperature distribution function is
		\begin{align}
			g_i(\mathbf{x}+\mathbf{e}_i\delta_t,t+\delta_t) - g_i(\mathbf{x},t) &= -(\mathbf{N}^{-1}\mathbf{Q})_{ij}\left[\mathbf{n}_j(x,t) - \mathbf{n}_j^{(\text{eq})}(x,t)\right] \notag \\
			&\quad (i=0,1,\cdots,q-1) 
			\label{eq:g}
		\end{align}
where $g_{i}$ is the temperature distribution function. 
For the D2Q5 lattice, $q = 5$; for the D3Q7 lattice, $q = 7$. 
The equilibrium moments $\mathbf{n}^{(\text{eq})}$ are given by, for the D2Q5 lattice
		\begin{equation}
				\mathbf{n}^{(\text{eq})}_{\text{D2Q5}} = \bigl[ T, \ uT, \ vT, \ a_{T}T, \ 0 \bigr]^T
				\label{eq:equilibriumMomentsD2Q5}
		\end{equation}
and for the D3Q7 lattice,
		\begin{equation}
				\mathbf{n}^{(\text{eq})}_{\text{D3Q7}} = \bigl[ T, \ uT, \ vT, \ wT, \ a_{T}T, \ 0, \ 0 \bigr]^T
				\label{eq:equilibriumMomentsD2Q5}
		\end{equation}
Here, $a_{T}$ is a parameter related to thermal diffusivity:  $a_{T}=20\sqrt{3}\alpha-4$ for D2Q5 lattice, and $a_{T}=42\sqrt{3}\alpha-6$ for D3Q7 lattice.
The macroscopic temperature ${T}$ is then obtained as $T=\sum_{i=0}^{q-1}g_i$.

In the standard LB model described above, it is implicitly assumed that the computational mesh is uniform and that the lattice spacing matches the distance traveled by the distribution functions in a single time step. 
Under this assumption, the evolution of the density distribution function (Eq. (\ref{eq:f})) can be decomposed into two sub-steps:
		\begin{flalign}
			&\text{Collision step: } f_i^+(\mathbf{x},t)=f_i(\mathbf{x},t)-\left(\mathbf{M}^{-1}\mathbf{S}\right)_{ij}\left[\mathbf{m}_j(\mathbf{x},t)-\mathbf{m}_j^{(\text{eq})}(\mathbf{x},t)\right]+\delta_tF_i^{\prime} &&  \label{eq:collisionF} \\
			&\text{Streaming step: } f_i(\mathbf{x}+\mathbf{e}_i\delta_t,t+\delta_t)=f_i^+(\mathbf{x},t) &&  \label{eq:streamingF}
		\end{flalign}
where $f_i^+(x,t)$ denotes the post-collision density distribution function. 
Similarly, the evaluation of the temperature distribution function (i.e., Eq. (\ref{eq:g})) is split into:
		\begin{flalign}
			&\text{Collision step: } g_i^+(\mathbf{x},t)=g_i(\mathbf{x},t)-(\mathbf{N}^{-1}\mathbf{Q})_{ij}\left[\mathbf{n}_j(\mathbf{x},t)-\mathbf{n}_j^{(\text{eq})}(\mathbf{x},t)\right] & \label{eq:collisionG}\\
			&\text{Streaming step: } g_i(\mathbf{x}+\mathbf{e}_i\delta_t,t+\delta_t)=g_i^+(\mathbf{x},t) & \label{eq:streamingG}
		\end{flalign}
where $g_i^+(\mathbf{x},t)$ represents the post-collision temperature distribution function. 
Further numerical details on the thermal LB method can be found in our previous works \cite{xu2017accelerated,xu2023multi,xu2019lattice}.

\subsection{The interpolation-supplemented LBM for simulating thermal convection on non-uniform meshes}

In this work, we adopt the procedure proposed by He et al. \cite{he1996some,he1997some,he1997error} to update the distribution function $f_{i}$ and $g_{i}$ on a non-uniform mesh. 
The key idea is to approximate the values of the distribution functions at grid points via interpolation from their corresponding values at off-lattice locations. 
Let $\mathbf{x}=(x_i,y_j)$ denote a grid point on an arbitrary rectangular non-uniform mesh in a Cartesian coordinate system. 
First, the collision steps (i.e. Eqs. (\ref{eq:collisionF}) and (\ref{eq:collisionG})) are performed locally at each grid point, yielding the post-collision distribution functions $f_i^+(\mathbf{x},t)$ and $g_i^+(\mathbf{x},t)$. 
Next, the virtual streaming steps (i.e. Eqs. (\ref{eq:streamingF}) and (\ref{eq:streamingG})) are applied to compute the distribution functions at the off-lattice locations $\mathbf{x}+\mathbf{e}_i\delta_t$, i.e., $f_i(\mathbf{x}+\mathbf{e}_i\delta_t,t+\delta_{t})$ and $g_i(\mathbf{x}+\mathbf{e}_i\delta_t,t+\delta_t)$. 
Unlike the uniform mesh case, where $\mathbf{x}+\mathbf{e}_i\delta_t$ coincides with a grid point, these locations are generally off-lattice in a non-uniform mesh. 
At such points, the values of $f_i(\mathbf{x}+\mathbf{e}_i\delta_t,t+\delta_t)$ and $g_i(\mathbf{x}+\mathbf{e}_i\delta_t,t+\delta_t)$ are set equal to the post-collision values $f_i^+(\mathbf{x},t)$ and $g_i^+(\mathbf{x},t)$, respectively. 
These off-lattice values are then interpolated back to the nearest grid point $\mathbf{x}$, providing the updated distribution functions $f_i(\mathbf{x},t+\delta_t)$ and $g_i(\mathbf{x},t+\delta_t)$. 
In this manner, the collision and the streaming steps are repeated iteratively. 
To preserve second-order spatial accuracy, a higher-order quadratic interpolation scheme following the formulation proposed by He \cite{he1997error} is employed.
This step is essential for consistent and rigorous recovery of the Navier-Stokes equations within the LBM framework.

\begin{figure}[htbp]
			\centering
			\includegraphics[width=1\linewidth]{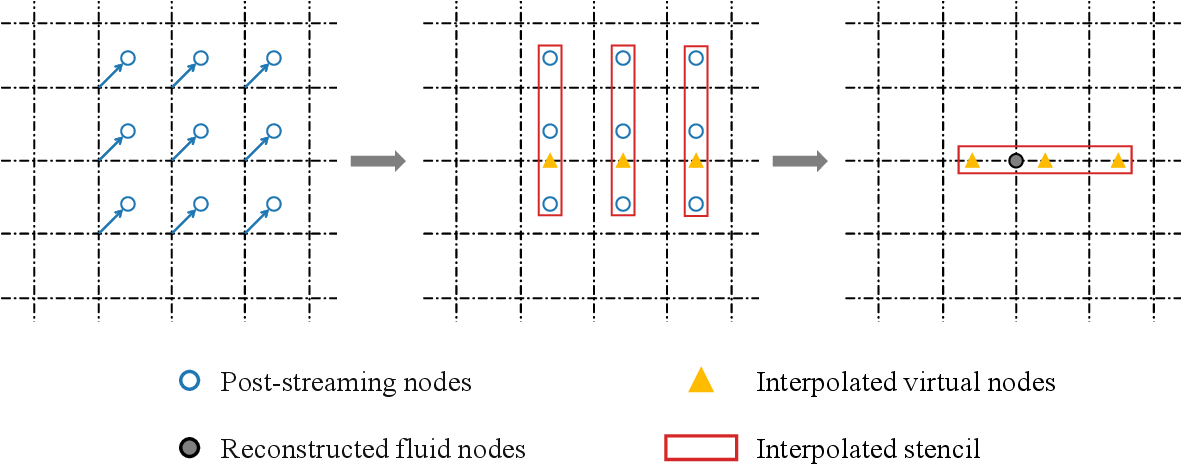}
			\caption{Illustration of the interpolation procedure in the interpolation-supplemented lattice Boltzmann method (ISLBM) on a two-dimensional non-uniform mesh.
}\label{fig_InterpolationDemo}
		\end{figure}

Fig. \ref{fig_InterpolationDemo} illustrates the interpolation procedure in the ISLBM for a two-dimensional non-uniform mesh. 
After the streaming step, the distribution functions generally arrive at off-lattice positions (see blue open circles). 
Quadratic interpolation stencils (see red boxes) are then constructed from neighboring post-streaming nodes to recover the distribution function at virtual nodes (see yellow triangles). 
The interpolated values at the virtual nodes are subsequently used to reconstruct the fluid nodes on the mesh (see black filled circles). 
This process completes one collision-streaming cycle on a non-uniform lattice.

In the interpolation kernel, the main contributions to the total computational cost are as follows. 
The first is off-lattice memory access, which requires fetching distribution values from non-contiguous global memory addresses, leading to irregular and non-coalesced reads. 
This constitutes the dominant bottleneck. 
The second is the evaluation of interpolation coefficients, which includes on-the-fly computation of quadratic interpolation weights (e.g., Lagrange polynomial coefficients) and depends on the degree of local mesh non-uniformity. 
The third is stencil selection and indexing, which involves identifying the appropriate neighboring nodes and determining their physical coordinates for accurate interpolation.

\subsection{Boundary conditions and implementation issues}

In the ISLBM framework, boundary conditions are implemented using the same kinetic boundary schemes as in uniform-mesh LBM. 
Specifically, the half-way bounce-back scheme is employed to enforce the no-slip velocity boundary condition, while the half-way anti-bounce-back scheme is applied for Dirichlet (constant-temperature) thermal boundaries and the half-way bounce-back scheme for Neumann (adiabatic) thermal boundaries.

At the fluid-solid interface, the no-slip velocity boundary condition is imposed using the half-way bounce-back scheme:
		\begin{equation}
			f_{\overline{i}}(\mathbf{x},t+\delta_t)=f_i^+(\mathbf{x},t) 
		\end{equation}
where, $f_{\overline{i}}(\mathbf{x},t)$ denotes the density distribution function corresponding to the discrete velocity $\mathbf{e}_{\overline{i}}=-\mathbf{e}_i$.

For thermal boundary conditions, the Dirichlet (constant-temperature) case is imposed via the half-way anti-bounce-back scheme:
		\begin{equation}
			g_{\bar{i}}(\mathbf{x},t+\delta_t)=-g_i^+(\mathbf{x},t)+\omega_{i} T_w 
		\end{equation}
where $T_{w}$ is the prescribed wall temperature. 
For the D2Q5 lattice, the weights are $\omega_{1-4}=(4+a_{T})/10$;
for the D3Q7 lattice, the weights are $\omega_{1-6}=(6+a_{T})/21$.
The Neumann (adiabatic) boundary condition is enforced using the half-way bounce-back scheme:
		\begin{equation}
			g_{\bar{i}}(\mathbf{x},t+\delta_t)=g_i^+(\mathbf{x},t)
		\end{equation}
where $g_{\overline{i}}(x,t)$ denotes the temperature distribution function associated with the discrete velocity $\mathbf{e}_{\bar{i}}$.

\begin{figure}[htbp]
			\centering
			\includegraphics[width=0.9\linewidth]{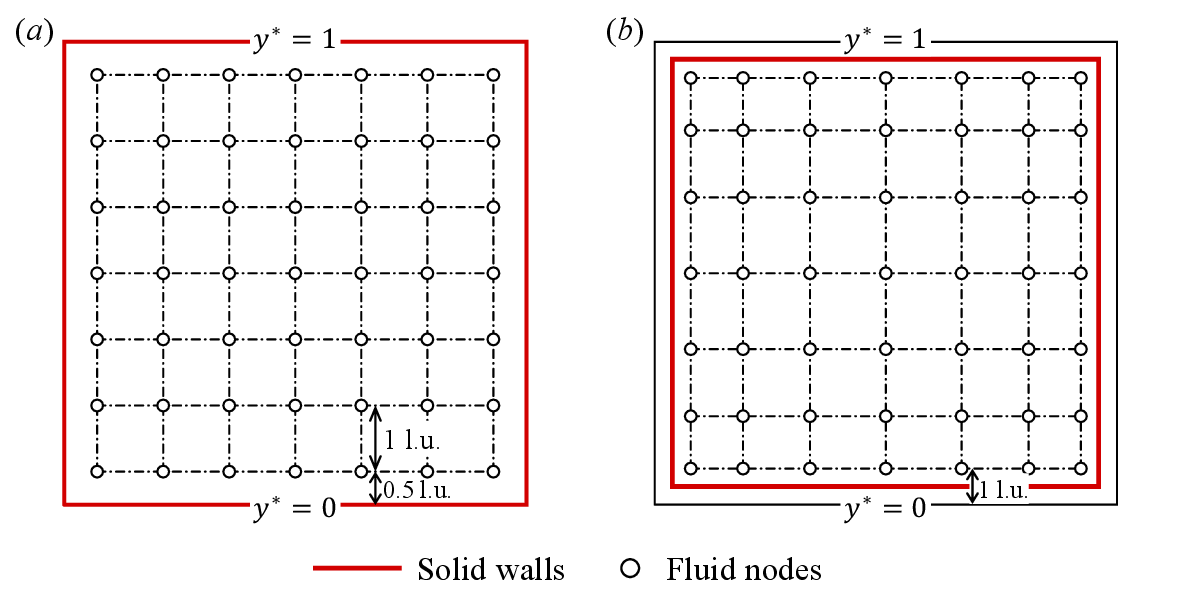}
			\caption{Examples of (\emph{a}) uniform mesh and (\emph{b}) non-uniform mesh with half-way bounce-back scheme. 
Here, l.u. denotes the lattice length-unit \cite{huang2015multiphasebook}. 
The first fluid node lies 0.5 l.u. from the wall due to the half-way bounce-back/anti-bounce-back scheme.}\label{fig_meshDemo}
		\end{figure}

We now discuss several implementation issues encountered when collecting data for post-analysis. 
The first concerns the distribution of grid points and the corresponding length scale. 
Assume convection is simulated within a cell of length $L$ and height $H$.
Fig. \ref{fig_meshDemo} illustrates typical examples of a uniform mesh and a non-uniform mesh with clustering near the walls. 
For a uniform mesh, the coordinates (taking the $y$-direction as an example) are specified as 

		\begin{equation}
			y_0^*=0,\quad y_{N_y+1}^*=1, 
			\label{eq:14a}
		\end{equation}
		\begin{equation}
			y_j^*=\frac{j-0.5}{N_y},\quad j=1,\cdots,N_y
			\label{eq:14b}
		\end{equation}
where the indices 1 and $N_{y}$ correspond to the fluid nodes nearest to the solid walls (denoted by the black circles in Fig. \ref{fig_meshDemo}a). 
The spatial resolution is uniform, with $\Delta y^*=1/N_y=1\mathrm{~l.u.}$, where "l.u." denotes the lattice length unit \cite{huang2015multiphasebook}.
Because the half-way bounce-back (or half-way anti-bounce-back) scheme is applied at the fluid-solid boundaries, the first layer of fluid nodes, $y_1^*$ and $y_{N_y}^*$, are offset by 0.5 l.u. from the solid walls (denoted by the red lines at $y_0^* = 0$ and $y_{N_y+1}^* = 1$).
This offset arises naturally from the half-way formulation and holds regardless of whether the mesh is uniform or non-uniform.

For non-uniform meshes, the coordinates (taking the $y$-direction as an example) are given by the error-function stretching \cite{Ceci2022}: 
		\begin{equation}
			y_j^*=\frac{1}{2}\left\{\frac{erf\left[a\left(\frac{j}{N_y+1}-0.5\right)\right]}{erf(a/2)}+1\right\},\quad j=0,1,\cdots,N_y+1
			\label{eq:erf}
		\end{equation}
where the indices 1 and $N_y$ correspond to the fluid nodes nearest to the solid walls, and the minimal spatial mesh spacing is $(\Delta y^*)_{min}=y_1^*=1$ l.u. 
Unlike the uniform case, the solid walls (denoted by the red lines in Fig. \ref{fig_meshDemo}b) are located at $y_{bottom}^*=0.5y_1^*$ and $y_{top}^*=1-0.5y_1^*$, so the effective cell height is $1-y_1^*$.
As marked in Fig. \ref{fig_meshDemo}(b), the offset between the boundary (red line) and the first fluid node (black circle) is 0.5 l.u. 
A summary of the length scales in the LB systems with uniform and non-uniform meshes is presented in Table \ref{tb:LBscale}.
			
		\begin{table}
			\centering
			\caption{Length scales in the physical system and in the LB systems with uniform and non-uniform meshes.}
			\label{tb:LBscale}
			\resizebox{0.99\textwidth}{!}{%
				\begin{tabular}{cccc}
					\toprule
					{} & Physical system & \makecell[l]{LB system with \\ uniform mesh} & \makecell[l]{LB system with \\ non-uniform mesh} \\
					\midrule
					Effective cell height & $H~\text{m}$ & $1$ & $1 - y_1^*$ \\
					Minimal resolution (1 l.u.) & -- & $1 / N_{y}$ & $y_1^*$ \\
					Characteristic length & -- & $N_y~\text{l.u.}$ & $(1 / y_1^* - 1)~\text{l.u.}$ \\
					\bottomrule
				\end{tabular}%
			}
		\end{table}

In the stretching error function (i.e. Eq. (\ref{eq:erf})), $a$ is a positive coefficient that controls the degree of stretching. 
In this work, we set $a = 1.5$ unless otherwise specified. 
A larger value of $a$ produces a more pronounced S-shape in the error function; consequently, mesh points are more densely packed near the boundaries and more sparsely distributed in the interior. 
Conversely, a smaller value of $a$ makes the error function vary more linearly across the domain. 
As $a$ approaches zero, the coordinate distribution $y_j^*$ approaches $j/N_{y+1}$, corresponding to a uniform mesh. 
To ensure the robustness of our mesh design, we quantitatively evaluated its quality using two widely accepted metrics. 
The first metric is the growth rate, defined as the ratio of adjacent cell sizes (e.g., $\Delta x_{i+1}/\Delta x_i$). 
Excessive growth rates can compromise numerical accuracy and stability. 
In our simulations with a $513 \times 513$ mesh and stretching parameter $a = 1.5$, the maximum growth rate is 1.0044, indicating a very smooth grid transition.
This value is well below the commonly accepted upper threshold of 1.2, ensuring high interpolation fidelity. 
The second metric is the aspect ratio, defined as the ratio of the longest to the shortest side of a computational cell. 
While elevated aspect ratios may be acceptable near boundary layers, where flow features are anisotropic, they are undesirable in core-flow regions. 
In our simulations, the maximum aspect ratio is 1.75, observed near the four boundaries. 
This value remains well within acceptable limits for structured grid-based simulations.

The second issue concerns the calculation of derivatives of flow variables at solid walls. 
The boundary schemes remain valid on non-uniform meshes, provided that the first fluid node adjacent to the wall is located 0.5 lattice units from the physical boundary. 
This half-spacing offset must be accounted for in derivative calculations. 
As shown in Fig. \ref{fig_meshDemo}(a), for a uniform mesh the first-order derivative of a flow variable $\phi$ at the wall is evaluated using a one-sided second-order finite-difference scheme:
		\begin{equation}
			\frac{\partial\phi}{\partial y}|_{bottom}=\frac{-8\phi_{wall}+9\phi_1-\phi_2}{3\Delta y},\quad\frac{\partial\phi}{\partial y}|_{top}=\frac{8\phi_{wall}-9\phi_{N_y}+\phi_{N_y-1}}{3\Delta y}
			\label{eq:16}
		\end{equation}
where $\Delta y=1$ l.u. is the minimal spatial resolution. 
For non-uniform meshes, the first-order derivative of $\phi$ at the wall is expressed as
		\begin{align}
			\left.\frac{\partial\phi}{\partial y}\right|_{\text{bottom}} &= \frac{-4q(q+1)\phi_{\text{wall}} + (2q+1)^2\phi_1 - \phi_2}{q(2q+1)\Delta y}, \nonumber \\
			\left.\frac{\partial\phi}{\partial y}\right|_{\text{top}}    &= \frac{4q(q+1)\phi_{\text{wall}} - (2q+1)^2\phi_N + \phi_{N-1}}{q(2q+1)\Delta y}. 
			\textbf{\label{eq:17}}
		\end{align}
where $q=\Delta y_2^*/\Delta y_1^*>1$ is the ratio of the second to the first mesh spacing, with $\Delta y_2^*=y_2^*-y_1^*$ and $\Delta y_1^*=y_1^*-y_0^*$.

The third issue is the calculation of spatial averages of flow variables. 
Based on the midpoint rule, the weighted average of a variable $\phi$ on a non-uniform mesh is computed as		
		\begin{equation}
			\langle\phi_j\rangle=\sum_{j=1}^{N_y}\phi_j\cdot\frac{y_{j+1}^*-y_{j-1}^*}{2}
			\label{eq:18}
		\end{equation}
This formulation ensures that the contribution of each data point to the overall average reflects its relative spatial extent within the domain. 
However, caution must be exercised when interpreting the results. 
As illustrated in Fig. \ref{fig_meshDemo}(b), the vertical extent of the flow domain does not span the full unit length, since the solid walls are located at $y_{bottom}^*=0.5y_1^*$ and $y_{top}^*=1-0.5y_1^*$, due to the half-way bounce-back scheme. 
Consequently, the average along the vertical flow direction should be normalized by the effective cell height $1-y_1^*$.

\section{Laminar convection in a 2-D side-heated cavity} \label{sec:SHC-2D}

We first consider thermal convection in the canonical 2-D side-heated cavity \cite{de1983natural}. 
The vertical walls are maintained at constant hot and cold temperatures, respectively, while the horizontal walls are adiabatic. 
All four walls enforce no-slip velocity boundary conditions. 
Although no exact analytical solution exists, this configuration has long served as a benchmark for coupled fluid flow and heat transfer, ever since Davis  \cite{de1983natural} reported comprehensive results on the flow structures and heat transport. 
A wide variety of solvers and datasets have been provided for convection at $Ra \leq 10^6$.
However, relatively fewer benchmark datasets are available at $Ra \geq 10^7$, since simulating thermal flows at higher Rayleigh number requires more robust numerical methods and greater computational resources. 
At high $Ra$, the thermal and velocity boundary layers become extremely thin, demanding high near-wall resolution. 
This is precisely where non-uniform meshes can deliver substantial computational savings without compromising accuracy, making the side-heated cavity a natural testbed for assessing both the accuracy and efficiency of ISLBM. 
Previously, we provided benchmark-quality results using a fine mesh of $2049^{2}$ uniform mesh points \cite{xu2017accelerated}, which were subsequently verified by Sun and Tao \cite{sun2023performance}, Chen et al. \cite{chen2022evolutions}, Vesper et al. \cite{Vesper2022}, and many others \cite{Ma2020,yang2021simulating,Zhao2021,Ma2022,Ma2022b,Xu2022,Lu2022,Nee2025,Feng2025}. 
Here, to validate the ISLBM for simulating coupled fluid flow and heat transfer, we present results for $10^6 \leq Ra \leq 10^8$ with the Prandtl number fixed at $Pr=0.71$. 
At these Rayleigh numbers, the flow remains steady and laminar in two dimensions, but the boundary layers become extremely thin, posing significant challenges for numerical resolution. 
Under these conditions, the flow is considered to have reached a steady state when the following convergence criteria are satisfied:
		\begin{equation}
			\begin{aligned}
				&\frac{\sum_i\|\mathbf{u}(x_i,t+2000\delta_t)-\mathbf{u}(x_i,t)\|_2}{\sum_i\|\mathbf{u}(x_i,t)\|_2} < 10^{-9}, \\
				&\frac{\sum_i|T(x_i,t+2000\delta_t)-T(x_i,t)|}{\sum_i|T(x_i,t)|} < 10^{-9}
			\end{aligned}
			\label{eq:steadyCriteria}
		\end{equation}
where $\left\|\mathbf{u}\right\|_2$ denotes $L^{2}$ norm of the velocity field $\mathbf{u}$.

In LBM simulations, a reference velocity must be specified to fully determine the relaxation parameters, owing to the weakly compressible nature of the method. 
For convection problems, we adopt the free-fall velocity $u_{f}=\sqrt{\beta g H \Delta_T}$ as the characteristic velocity, which yields the dimensionless Mach number $Ma=u_f/c_s=\sqrt{\beta gH\Delta_T}/c_s$.
Here, $c_s=1/\sqrt{3}$ l.u./t.s. is the lattice speed of sound.
Unless otherwise stated, we set $Ma=0.1$ in all simulations. 
To approximate incompressibility, the Mach number should be kept as small as possible. 
In LBM simulations, however, the ratio of the dimensionless time step to the dimensionless spatial resolution is given by $dt/dx=(\delta_t/\sqrt{H/(\beta g\Delta_T)})/(\delta_x/H)=\sqrt{\beta gH\Delta_T}=Ma\cdot c_s$, with $\delta_{x}=1$ l.u. and $\delta_t=1$ t.s. 
Thus, reducing the Mach number decreases the dimensionless time step for a fixed mesh resolution, which in turn reduces computational efficiency in time marching.
Therefore, our choice of $Ma=0.1$ represents a compromise between approximating incompressibility and maintaining computational efficiency.

\begin{figure}[tbp]
			\centering
			\includegraphics[width=1\linewidth]{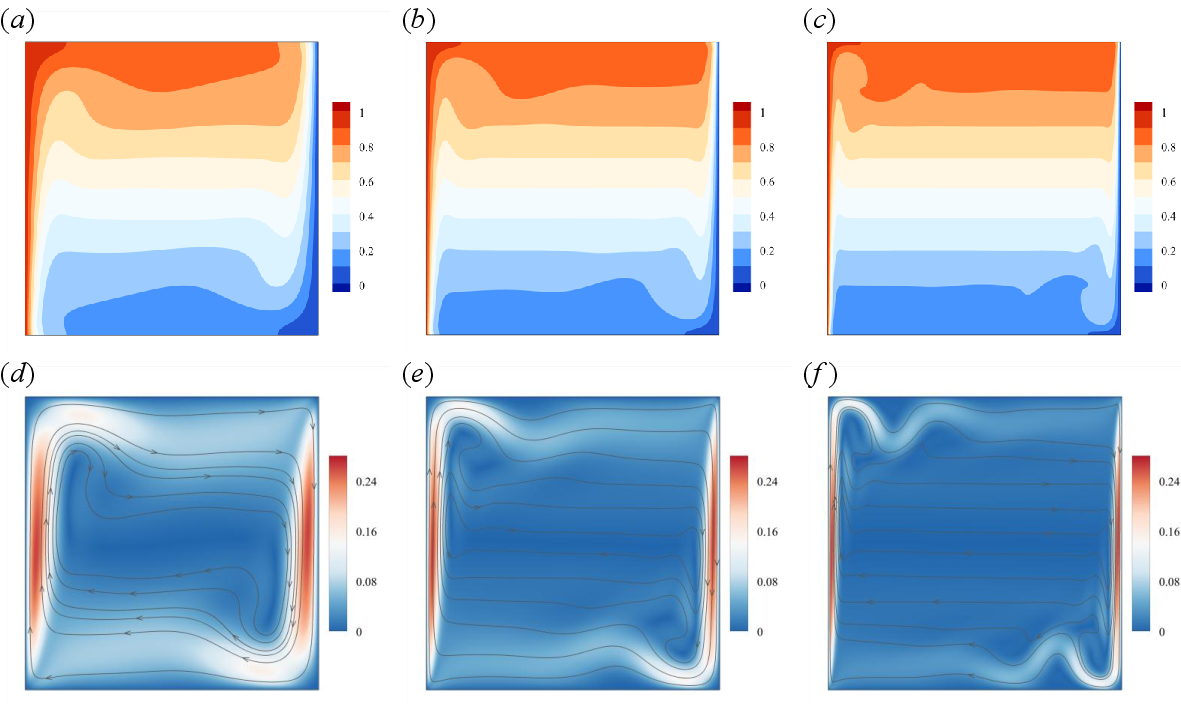}
			\caption{Contours of (\emph{a}-\emph{c}) temperature field $T^{*}$ and (\emph{d}-\emph{f}) velocity magnitude $\sqrt{u^{*2}+v^{*2}}$ (superposed with streamlines), at (\emph{a}, \emph{d}) $Ra=10^6$, (\emph{b}, \emph{e}) $Ra=10^7$, and (\emph{c}, \emph{f}) $Ra=10^8$.}\label{fig3_temperatureVelocity2D}
		\end{figure}

In Fig. \ref{fig3_temperatureVelocity2D}, we present the temperature and velocity fields. 
At $Ra=10^6$ (see Fig. \ref{fig3_temperatureVelocity2D}a), the temperature distribution exhibits relatively thick thermal boundary layers along the hot (left) and cold (right) walls, with smooth stratification in the interior. 
As $Ra$ increases to $10^{7}$ and $10^{8}$ (see Figs. \ref{fig3_temperatureVelocity2D}b and \ref{fig3_temperatureVelocity2D}c), these boundary layers become thinner, reflecting enhanced thermal gradients and stronger convective transport. 
The velocity fields reveal a single dominant clockwise large-scale circulation cell in all cases (see Figs. \ref{fig3_temperatureVelocity2D}d-f), with elevated velocity magnitudes near the boundary layers. 
At higher $Ra$, the maximum velocities are concentrated in narrow shear layers adjacent to the sidewalls, while the streamlines indicate intensified plume activity near the vertical boundaries. 
The core region remains relatively quiescent and horizontally stratified. 
Overall, the flow structure agrees well with earlier 2-D simulations (see Fig. 10 in our previous work \cite{xu2017accelerated} and with many other studies \cite{sun2023performance,chen2022evolutions,Vesper2022}).
 
\begin{figure}[htbp]
			\centering
			\includegraphics[width=0.9\linewidth]{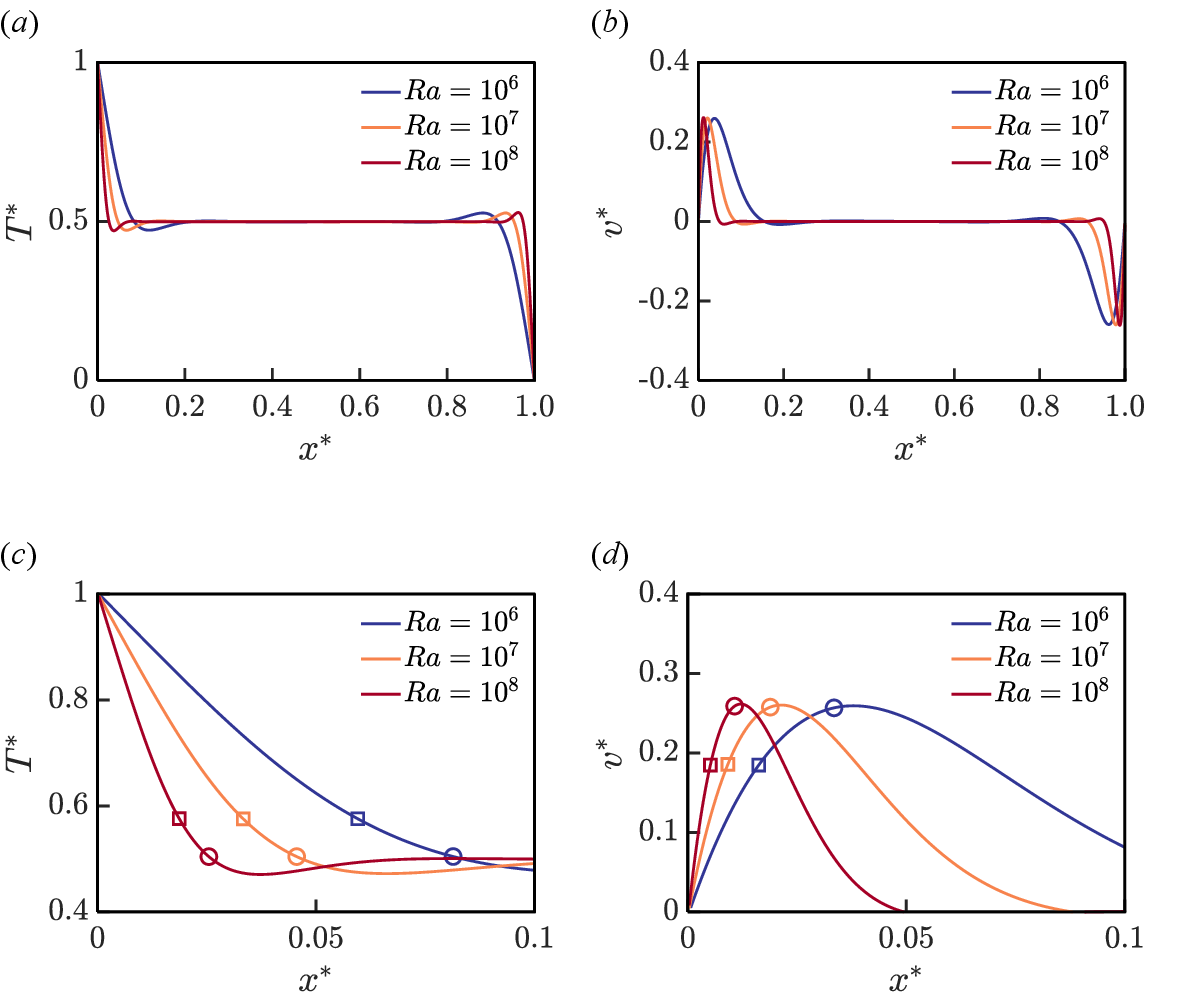}
			\caption{Horizontal distributions of vertically averaged (\emph{a}) temperature $T^{*}$ and (\emph{b}) vertical velocity $v^{*}$.
				Panels (\emph{c}, \emph{d}) show enlarged views near the hot wall $(x^{*}=0)$, 
				where open circles denote boundary layer locations determined by the 99\% criterion method and open squares denote those determined by the slope method.}\label{fig_verticalAvg}
		\end{figure}

Figs. \ref{fig_verticalAvg}(a) and \ref{fig_verticalAvg}(b) show the horizontal distributions of the temperature $T^{*}$ and vertical velocity $v^{*}$, each averaged over the mid-plane band $0.4 \leq y \leq 0.6$ for three Rayleigh numbers. 
This averaging minimizes end-wall effects and yields smooth profiles suitable for boundary-layer analysis.
Pronounced gradients are observed near the sidewalls, reflecting the development of thermal and momentum boundary layers adjacent to the hot and cold boundaries. 
To examine these boundary layers in more detail, Figs. \ref{fig_verticalAvg}(c) and \ref{fig_verticalAvg}(d) present enlarged views near the hot wall $(x^*=0)$. 
The boundary-layer thicknesses are determined using two widely adopted approaches: the 99\% criterion method and the slope method.
In the 99\% criterion method, the thermal boundary layer thickness $\delta_{T}$ is defined as the distance from the wall where $T^{*}$ attains 99\% of the difference between the wall and the core value, while the velocity boundary layer thickness $\delta_{V}$  is defined as the distance where $v^{*}$ reaches 99\% of its maximum magnitude. 
In the slope method, $\delta_{T}$ is obtained as the ratio of the core-to-wall temperature difference to the near-wall temperature gradient, and $\delta_{V}$  as the ratio of the maximum velocity to the near-wall velocity gradient. 
Power-law  fits show that both methods yield consistent results, with $\delta_{T} \sim Ra^{-0.25}$ and  $\delta_{V} \sim Ra^{-0.25}$ at $Pr=0.71$. 
These results confirm that the thermal and velocity boundary layers exhibit comparable scaling with Rayleigh number.

In Table \ref{tb:2DNuRe}, we present quantitative results for heat transfer efficiency in terms of the Nusselt number, and for global flow strength in terms of the Reynolds number. 
Specifically, the volume-averaged Nusselt number $\langle{Nu}\rangle$ is defined as 
		\begin{equation}
			\langle Nu \rangle=\sqrt{RaPr}\langle u^*T^*\rangle_V+1
		\end{equation}
the average Nusselt number along the hot wall $Nu_{\text{hot}}$ is given by
		\begin{equation}
			Nu_{\text{hot}}=-\langle\frac{\partial T^*}{\partial x^*}\rangle_{x^*=0}
		\end{equation}
and the average Nusselt number along the vertical centerline $Nu_{\text{middle}}$ is expressed as
		\begin{equation}
			Nu_{\text{middle}}=\langle\sqrt{Ra Pr}u^*T^*-\frac{\partial T^*}{\partial x^*}\rangle_{x^*=0.5}
		\end{equation}
The volume-averaged Reynolds number $\langle Re \rangle$ is defined as
		\begin{equation}
			\langle Re \rangle=\sqrt{\frac{Ra}{Pr}}\langle\sqrt{u^{*2}+v^{*2}}\rangle_V
		\end{equation}
Here, $\langle\cdot\rangle_{V}$ denotes a volume average, $\langle\cdot\rangle_{x^*=0}$ a line average along the hot wall, and $\langle\cdot\rangle_{x^*=0.5}$ a line average along the vertical centerline. 
For comparison, we also include benchmark data from previous studies: 
Wang et al. \cite{wang2019non}, who employed a finite-difference method to solve low-Mach-number equations; 
Le Quéré \cite{le1991accurate}, who used a pseudo-spectral Chebyshev algorithm; 
Contrino et al. \cite{contrino2014lattice}, who applied an MRT-LBM with a forcing term split using the Strang--Marchuk scheme; 
and our previous work \cite{xu2017accelerated}, which used the LBM with Guo’s forcing scheme \cite{guo2002discrete,guo2008analysis,chai2012effect} on a uniform mesh.
		\begin{table}[htbp]
			\centering
			\caption{Benchmark solutions of the Nusselt numbers and Reynolds number. Columns from left to right indicate: 
Rayleigh number $Ra$; 
flow database; 
grid number $N_x\times N_y$; 
volume-averaged Nusselt number $\langle Nu \rangle$; 
average Nusselt number along the hot wall $Nu_{\text{hot}}$; 
average Nusselt number along the vertical centerline $Nu_{\text{middle}}$; volume-averaged Reynolds number $\langle Re \rangle$.}
			\resizebox{1\textwidth}{!}{
				\begin{tabular}{ccccccc}
					\toprule
					\makecell[c]{$Ra$} & \makecell[c]{Flow database} & \makecell[c]{$N_x \times N_y$} & \makecell[c]{$\langle Nu \rangle$} & \makecell[c]{$Nu_{\text{hot}}$} & \makecell[c]{$Nu_{\text{middle}}$} & \makecell[c]{$\langle Re \rangle$} \\
					\midrule
					\multirow[t]{8}{*}{$10^6$} 
					& Present & $257^2$  & 8.8282 & 8.8509 & 8.8277 & 99.6110 \\
					& Present & $513^2$  & 8.8202 & 8.8317 & 8.8200 & 99.3839 \\
					& Present & $1025^2$ & 8.8188 & 8.8246 & 8.8188 & 99.3576 \\
					& $\infty$ & --      & 8.8186 & 8.8206 & 8.8186 & 99.3542 \\
					& $p$      & --      & 2.6    & 1.5    & 2.6    & 3.1 \\
                    & GCI (\%) & --      & 0.0037 & 0.058  & 0.0033 & 0.0043 \\ [5pt]   
					& Wang \cite{wang2019non} & $128^2$ & 8.830 & -- & -- & 99.11 \\
					& Le Quéré \cite{le1991accurate} & $72^2$ & -- & 8.825 & 8.825 & -- \\
					& Contrino \cite{contrino2014lattice} & $2043^2$ & 8.8252 & 8.8252 & 8.8252 & -- \\[1pt] 
				    \hline \\[1pt]

					\multirow[t]{8}{*}{$10^7$} 
					& Present & $513^2$  & 16.5245 & 16.5461 & 16.5243 & 230.6601 \\
					& Present & $1025^2$ & 16.5133 & 16.5242 & 16.5133 & 230.2448 \\
					& Present & $2049^2$ & 16.5114 & 16.5169 & 16.5114 & 230.1948 \\
					& $\infty$ & --      & 16.5110 & 16.5132 & 16.5110 & 230.1879 \\
					& $p$      & --      & 2.5     & 1.6     & 2.6     & 3.1 \\ 
                    & GCI (\%) & --      & 0.0030  & 0.028   & 0.0029  & 0.0037 \\ [5pt]   
					& Xu \cite{xu2017accelerated} & $2049^2$ & 16.52414 & 16.52229 & 16.52428 & -- \\
					& Wang \cite{wang2019non} & $256^2$ & 16.528 & -- & -- & 229.70 \\
					& Le Quéré \cite{le1991accurate} & $80^2$ & -- & 16.523 & 16.523 & -- \\
					& Contrino \cite{contrino2014lattice} & $2043^2$ & 16.5231 & 16.5233 & 16.5232 & -- \\[1pt]  
				    \hline \\[1pt]
					
					\multirow[t]{8}{*}{$10^8$} 
					& Present & $513^2$  & 30.3130 & 30.3503 & 30.3125 & 542.9352 \\
					& Present & $1025^2$ & 30.2225 & 30.2423 & 30.2223 & 537.3420 \\
					& Present & $2049^2$ & 30.2072 & 30.2172 & 30.2072 & 536.6449 \\
					& $\infty$ & --      & 30.2041 & 30.2096 & 30.2041 & 536.5456 \\
					& $p$      & --      & 2.6     & 2.1     & 2.6     & 3.0 \\ 
                    & GCI (\%) & --      & 0.013   & 0.031   & 0.013   & 0.023 \\ [5pt]   
					& Xu \cite{xu2017accelerated} & $2049^2$ & 30.23013 & 30.22650 & 30.23050 & -- \\
					& Wang \cite{wang2019non} & $256^2$ & 30.242 & -- & -- & 534.23 \\
					& Le Quéré \cite{le1991accurate} & $128^2$ & -- & 30.225 & 30.225 & -- \\
					& Contrino \cite{contrino2014lattice} & $2043^2$ & 30.2255 & 30.2268 & 30.2261 & -- \\
					\bottomrule
				\end{tabular} \label{tb:2DNuRe}
			}
		\end{table}
		
The comparison in Table \ref{tb:2DNuRe} shows that the present data converge toward the asymptotic value (discussed later) and are in good agreement with previous studies. 
An interesting observation is that the present Nusselt numbers are slightly lower than those reported in our earlier work \cite{xu2017accelerated}, with deviations of less than 0.1\%.
We attribute this small discrepancy to the weak compressibility of the current LB model. 
In our previous work \cite{xu2017accelerated}, only density fluctuation $\delta\rho$ were included in certain components of the equilibrium moments $\mathbf{m}^{(\mathrm{eq})}$ (see Eq. (3) in \cite{xu2017accelerated}), which provided a good approximation for incompressible flows in steady state. 
In the present study, however, we incorporate the full density $\rho$ in the equilibrium moments $\mathbf{m}^{(\text{eq})}$, resulting in a weakly compressible formulation that enhances numerical stability. 
As also reported by Wang et al. \cite{wang2019non} and Wen et al. \cite{Wen2024,Wen2025}, weak compressibility leads to a slight decrease in $Nu$ and a slight increase in $Re$, without significantly affecting the scaling exponents with respect to $Ra$.

\begin{figure}[htbp]
			\centering
			\includegraphics[width=1\linewidth]{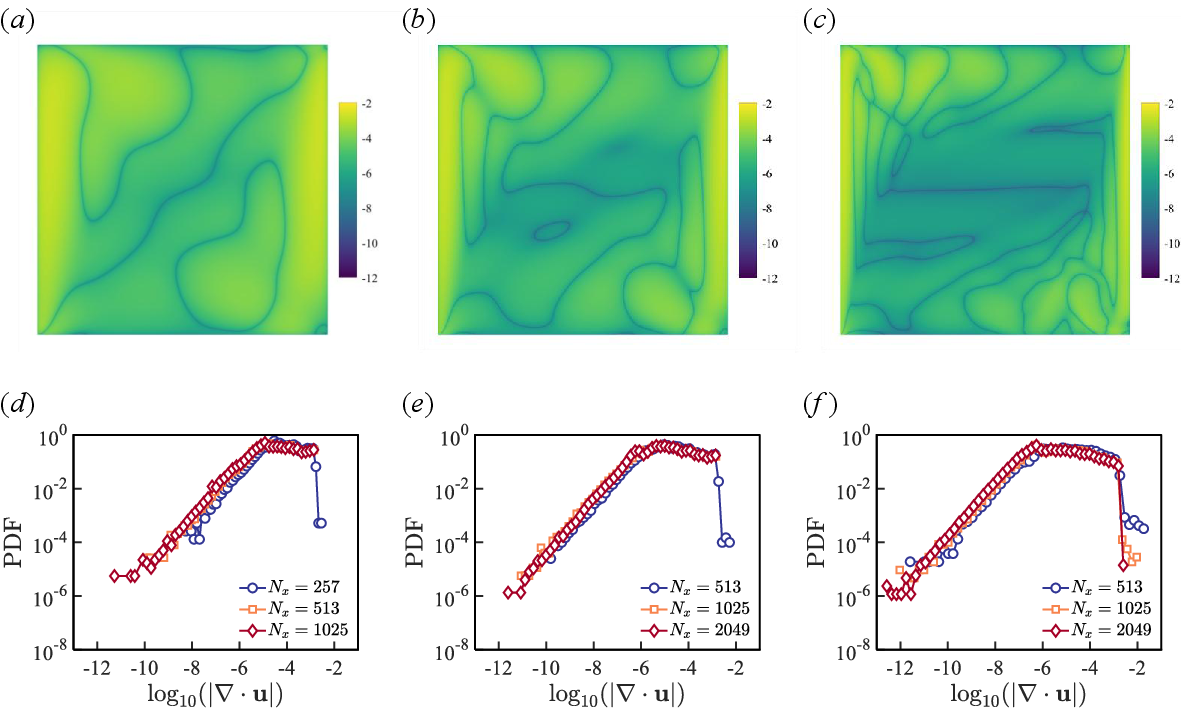}
			\caption{(\emph{a}-\emph{c}) Contours of the logarithm of velocity divergence magnitude, $\log_{10}(|\nabla\cdot\mathbf{u}|)$.
(\emph{d}-\emph{f}) Probability density functions (PDFs) of $\log_{10}(|\nabla\cdot\mathbf{u}|)$, obtained over the entire cell, at (\emph{a}, \emph{d}) $Ra = 10^6$, (\emph{b}, \emph{e}) $Ra = 10^7$ and (\emph{c}, \emph{f}) $Ra = 10^8$.}\label{fig_divU2D}
		\end{figure}

To assess the compressibility effects intrinsic to the LBM employed in this study, we examine both the spatial distribution and statistical properties of the velocity divergence field. 
Although the LBM formulation is nearly incompressible, the simulations operate in a weakly compressible regime with a Mach number of $Ma=0.1$, necessitating a careful evaluation of divergence behavior. 
Figs. \ref{fig_divU2D}(a-c) show contour maps of the base-10 logarithm of the velocity divergence magnitude, $\log_{10}(|\nabla\cdot\mathbf{u}|)$, computed over the entire domain. 
In all cases, the divergence field remains predominantly low, with most of the domain exhibiting values below $10^{-4}$, consistent with the nearly incompressible character of the formulation. 
At $Ra=10^6$ (see Fig. \ref{fig_divU2D}a), the divergence field is characterized by smoother structures, reflecting weak localized deviations from incompressibility. 
As $Ra$ increases to $10^7$ (see Fig. \ref{fig_divU2D}b), finer features emerge, with localized regions of elevated divergence magnitude appearing intermittently throughout the domain. 
At the highest Rayleigh number considered, $Ra=10^8$ (see Fig. \ref{fig_divU2D}c), the divergence field develops elongated and filamentary structures aligned with regions of intensified thermal plumes and sharp velocity gradients. 
Although local compressibility effects are slightly amplified at higher Rayleigh numbers, they remain small in magnitude and spatially confined.
This ensures that the weakly compressible LBM faithfully reproduces the incompressible flow characteristics.

To quantitatively assess the divergence statistics, Figs. \ref{fig_divU2D}(d-f) show the probability density functions (PDFs) of $\log_{10}(|\nabla\cdot\mathbf{u}|)$, evaluated at three spatial resolutions for each Rayleigh numbers. 
In all cases, the PDFs peak within $|\nabla\cdot\mathbf{u}|\in[10^{-7},10^{-3}]$, with long tails extending down to $10^{-11}$, again consistent with a nearly incompressible flow field. 
As resolution increases, the PDFs converge, particularly near the most probable values. 
In addition, a modest increase in the domain-averaged divergence is observed when $Ra$ increases from $10^6$ to $10^8$. 
This trend is physically reasonable, as stronger thermal forcing steepens velocity and temperature gradients, thereby amplifying local compressibility effects in weakly compressible formulations. 
Similar divergence characteristics have also been reported in high-fidelity incompressible simulations. 
For example, Ostilla-Mónico et al. \cite{Ostilla2015a}  observed locally non-solenoidal fields with residual divergence of order of $O(10^{-3})$ in multiple-resolution incompressible Navier--Stokes simulations.
To the best of our knowledge, such low levels of divergence do not introduce significant qualitative or quantitative deviations, even in turbulent convection regimes.

Our non-uniform mesh is systematically refined to preserve geometric similarity while halving the effective mesh spacing at each refinement level. 
To assess convergence, we employ Richardson extrapolation using three successively refined meshes. 
For $Ra=10^6$, the grid resolutions are $(N_{\text{coarse}},N_{\text{medium}},N_{\text{fine}})=(257,513,1025)$, and for $Ra=10^7$ and $10^8$, they are $(N_{\text{coarse}},N_{\text{medium}},N_{\text{fine}})=(513,1025,2049)$. 
Throughout, the effective mesh spacing follows $\Delta x_i\propto1/N_i$, with refinement ratio $r=\Delta x_{\text{coarse}}/\Delta x_{\text{medium}}=\Delta x_{\text{medium}}/\Delta x_{\text{fine}}=2$. 
Let $\mathcal{F}(\Delta x)$ denote a flow quantity computed on mesh spacing $\Delta x$. 
The observed order of convergence $p$ is then estimated as
		\begin{equation}
		p=\frac{\ln\left(\left|\frac{\mathcal{F}(\Delta x_{\text{coarse}})-\mathcal{F}(\Delta x_{\text{medium}})}{\mathcal{F}(\Delta x_{\text{medium}})-\mathcal{F}(\Delta x_{\text{fine}})}\right|\right)}{\ln(r)}
		\end{equation}
The corresponding asymptotic, grid-independent value is obtained as
		\begin{equation}
			\mathcal{F}(1/\infty)\approx\frac{r^p\mathcal{F}(\Delta x_{\text{fine}})-\mathcal{F}(\Delta x_{\text{medium}})}{r^p-1}
		\end{equation}
Table \ref{tb:2DNuRe} reports the estimated convergence orders and extrapolated asymptotic values of the Nusselt and Reynolds numbers. 
Although the table lists four significant digits for readability, all computations were performed with at least fourteen-digit precision. 
The interpolation-supplemented LBM solver achieves nearly third-order spatial accuracy in domain-integrated quantities, such as volume-averaged Nusselt and Reynolds numbers.
This enhanced accuracy arises because the integration process inherently damps local discretization errors.
To further examine this hypothesis, Table \ref{tb:2Dvut} reports convergence orders for local flow variables evaluated at the cell center.
The vorticity field, being a derivative of velocity, exhibits close to third-order accuracy, while local temperature and velocity converge at approximately second-order accuracy.

In addition to assessing the observed order of convergence and extrapolated asymptotic values, we further quantified the discretization uncertainty using the Grid Convergence Index (GCI), following the ASME procedure \citep{Celik2008}. 
For three systematically refined meshes with a refinement ratio of $r$, the GCI on the fine mesh is computed as
\begin{equation}
\text{GCI}_{\text{fine}} =
\frac{F_s}{r^p - 1}
\left|\frac{\mathcal{F}(\Delta x_{\text{fine}}) - \mathcal{F}(\Delta x_{\text{medium}})}{\mathcal{F}(\Delta x_{\text{fine}})}\right|
\times 100\% ,
\end{equation}
where $p$ is the observed convergence order and $F_s=1.25$ is the safety factor recommended for three-grid studies. 
The resulting GCI values are typically below $0.1\%$ for global quantities such as the Nusselt and Reynolds numbers (see Table \ref{tb:2DNuRe}), whereas local velocity components exhibit somewhat larger values, up to $\sim 2\%$ (see Table \ref{tb:2Dvut}), reflecting their higher sensitivity to grid resolution.
These results demonstrate that the fine-mesh solutions lie within the asymptotic range of convergence and that the numerical uncertainty in the reported quantities is negligible.

		\begin{table}[htbp]
			\centering
			\caption{Convergence behavior of flow variables at cell center. The columns from left to right indicate the following: Rayleigh number $Ra$; absolute value of vorticity $|\omega_c|$; temperature $T_c$, horizontal velocity component $u_c$; vertical velocity component $v_c$.}
			\label{tb:2Dvut}
			\resizebox{1\textwidth}{!}{
				\begin{tabular}{cccccc}
					\toprule
					\makecell[c]{$Ra$} &
					\makecell[c]{Grid} &
					\makecell[c]{$|\omega_c|$} &
					\makecell[c]{$T_c$} &
					\makecell[c]{$u_c$} &
					\makecell[c]{$v_c$} \\
					\midrule
					\multirow[t]{5}{*}{$10^6$} 
					& $257^2$  & 0.1175 & 0.49945 & -1.2163$\times10^{-4}$ & 1.8504$\times10^{-5}$ \\
					& $513^2$  & 0.1158 & 0.49947 & -1.3013$\times10^{-4}$ & 1.9377$\times10^{-5}$ \\
					& $1025^2$ & 0.1156 & 0.49948 & -1.3239$\times10^{-4}$ & 1.9551$\times10^{-5}$ \\
					& $\infty$ & 0.1155 & 0.49948 & -1.3320$\times10^{-4}$ & 1.9594$\times10^{-5}$ \\
					& $p$      & 2.6    & 2.0     & 1.9                    & 2.3 \\
					& GCI (\%) & 0.054  & 0.00055 & 0.77                   & 0.27 \\
                    \hline \\ [1pt]
                    
					\multirow[t]{5}{*}{$10^7$} 
					& $513^2$  & 0.1088 & 0.49930 & -8.0524$\times10^{-5}$ & 8.2430$\times10^{-6}$ \\
					& $1025^2$ & 0.1076 & 0.49934 & -8.3466$\times10^{-5}$ & 8.3163$\times10^{-6}$ \\
					& $2049^2$ & 0.1075 & 0.49934 & -8.4272$\times10^{-5}$ & 8.3330$\times10^{-6}$ \\
					& $\infty$ & 0.1074 & 0.49935 & -8.4576$\times10^{-5}$ & 8.3379$\times10^{-6}$ \\
					& $p$      & 2.7    & 2.0     & 1.9                    & 2.1 \\
					& GCI (\%) & 0.037  & 0.00068 & 0.45                   & 0.073 \\
					\hline \\ [1pt]

					\multirow[t]{5}{*}{$10^8$} 
					& $513^2$  & 4.1046$\times10^{-2}$ & 0.49911 & -3.7830$\times10^{-5}$ & 3.1377$\times10^{-6}$ \\
					& $1025^2$ & 3.9075$\times10^{-2}$ & 0.49922 & -4.2857$\times10^{-5}$ & 2.7224$\times10^{-6}$ \\
					& $2049^2$ & 3.8864$\times10^{-2}$ & 0.49925 & -4.4429$\times10^{-5}$ & 2.6433$\times10^{-6}$ \\
					& $\infty$ & 3.8838$\times10^{-2}$ & 0.49926 & -4.5144$\times10^{-5}$ & 2.6248$\times10^{-6}$ \\
					& $p$      & 3.2                   & 2.0     & 1.7                    & 2.4 \\
					& GCI (\%) & 0.082                 & 0.0024  & 2.0                    & 0.88 \\
					\bottomrule
				\end{tabular}%
			}
		\end{table}
		
We evaluate the computational efficiency of our in-house LBM solver by comparing it against two widely used open-source solvers: 
Nek5000 (version v19.0), which is based on the spectral element method (SEM) \cite{Fischer1997}, and OpenFOAM (version 8), which employs the finite volume method (FVM) \cite{Weller1998}.

The Nek5000 solver extends the standard finite element method to higher-order polynomial basis functions \cite{Fischer1997}. 
In this study, the polynomial order for both velocity and pressure is set to $N=8$, following the Pn–Pn formulation. 
This choice is consistent with values commonly adopted in high-fidelity turbulent flow simulations. 
The Pn–Pn formulation employs a time-splitting scheme that decouples the pressure and velocity fields into three distinct substeps. 
The pressure Poisson equation is solved using the preconditioned Generalized Minimal Residual (GMRES) method, while the Helmholtz equations for the velocity components are solved using the preconditioned Conjugate Gradient (CG) method. 
For temporal discretization, a semi-implicit scheme is employed. 
The viscous term is discretized implicitly using a second-order backward difference scheme, while the nonlinear convection term is discretized explicitly using a second-order extrapolation scheme. 
The energy equation, governing temperature via a convection-diffusion formulation, is handled analogously: 
the transient and diffusion terms are treated implicitly with a second-order backward difference scheme, while the convection term is treated explicitly using a second-order extrapolation scheme.

The OpenFOAM solver enforces the integral form of the conservation equations on each control volume. 
In this study, we employed the transient \emph{buoyantPimpleFoam} solver \cite{Weller1998}. 
The Navier-Stokes equations were solved using the PIMPLE pressure-velocity coupling algorithm. 
Within the inner iterations, the Semi-Implicit Method for Pressure-Linked Equations (SIMPLE) scheme was applied for pressure correction, while the outer iterations used the Pressure Implicit with Splitting of Operators (PISO) scheme to handle transient terms, thereby improving convergence and numerical stability. 
For temporal discretization, a second-order implicit backward scheme was adopted. 
Spatial discretization was carried out within the Gauss linear framework, in which volume integrals are evaluated by Gaussian quadrature and face fluxes are approximated using linear interpolation.
The pressure equation in the resulting linear systems was solved using the Geometric--Algebraic Multi-Grid (GAMG) method to accelerate convergence for large-scale problems. 
The velocity and energy equations were solved using the Preconditioned Bi-Conjugate Gradient Stabilized (PBiCGStab) method, which efficiently handles non-symmetric sparse matrices while maintaining both stability and accuracy.

\begin{figure}[htbp]
			\centering
			\includegraphics[width=0.9\linewidth]{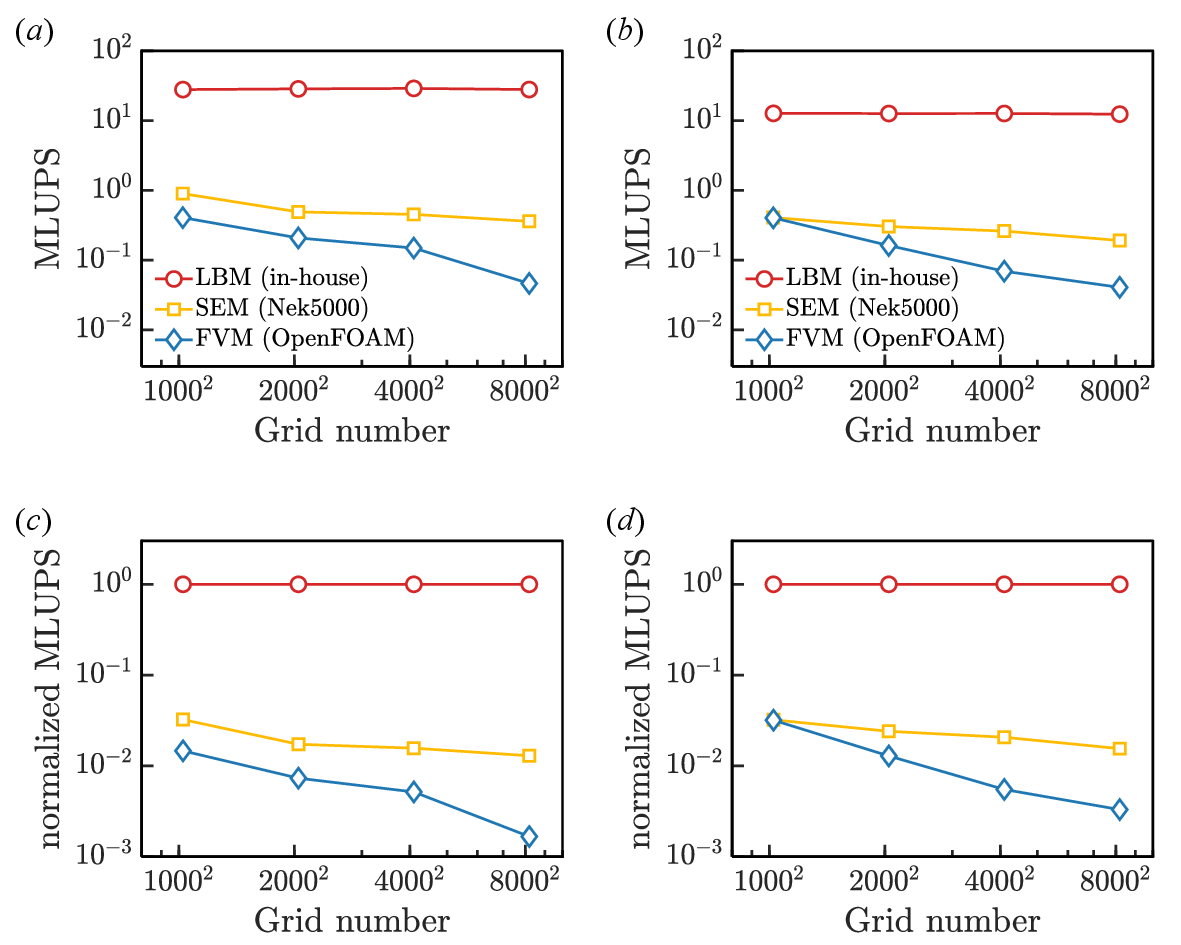}
			\caption{Performance comparison of three solvers in terms of million lattice updates per second (MLUPS): an in-house solver using the lattice Boltzmann method (LBM), Nek5000 using the spectral element method (SEM), and OpenFOAM using the finite volume method (FVM).
(\emph{a}, \emph{b}) Absolute MLUPS as a function of grid number; (\emph{c}, \emph{d}) MLUPS normalized by the corresponding LBM values, on (\emph{a}, \emph{c}) uniform and (\emph{b}, \emph{d}) non-uniform meshes.}\label{fig_compare_MLUPS_2D}
		\end{figure}

As a performance metric, we adopt the million lattice updates per second (MLUPS), defined as:
		\begin{equation}
			\mathrm{MLUPS} = \frac{\text{grid size} \times \text{iteration steps}}{\text{running time} \times 10^6} 
			\label{eq:26}
		\end{equation}
A higher MLUPS value indicates greater computational efficiency. 
The use of simple geometry and boundary conditions in the side-heated cavity enables direct and fair comparisons with Nek5000 and OpenFOAM, while minimizing discrepancies related to solver-specific geometrical handling or mesh preprocessing.
This ensures that the reported performance metrics primarily reflect the core algorithmic efficiency, rather than differences in geometry setup. 
Figs. \ref{fig_compare_MLUPS_2D}(a) and \ref{fig_compare_MLUPS_2D}(b) show MLUPS as a function of grid number on uniform and non-uniform meshes, respectively.
The simulations were carried at $Ra=10^8$ and $Pr=0.71$, on AMD EPYC 9135 CPUs. 
For non-uniform meshes, the stretching coefficient in the error-function (i.e. Eq. (\ref{eq:erf})) was set to $a=2.1$. 
Increasing the grid resolution from $1024^2$ to $8192^2$ has minimal impact on the performance of the in-house LBM solver. 
In contrast, Nek5000 shows a slight decline, while OpenFOAM exhibits a significant drop in efficiency. 
To further emphasize the relative differences, Figs. \ref{fig_compare_MLUPS_2D}(c) and \ref{fig_compare_MLUPS_2D}(d) present MLUPS values normalized by those of the in-house LBM solver at each grid size. 
The results clearly show that Nek5000 becomes up to two orders of magnitude slower than LBM at the largest grid size of $8192^2$. 
For OpenFOAM, the disparity is even more pronounced, reaching up to three orders of magnitude slower at the same resolution. 
The observed performance advantage of the LBM  stems from its reliance almost entirely on local operations (collision and streaming) with no global matrix solves or complex stencil dependencies. 
This design yields excellent memory locality and computational regularity, allowing near-ideal scaling across CPU cores until memory-bandwidth limits are reached. 
In contrast, SEM and FVM require global or semi-global solves for the pressure field and other coupled variables, involving sparse matrix–vector products, iterative solvers, and halo exchanges. 
These introduce higher communication costs and memory-access irregularity. 
Consequently, the orders-of-magnitude speedup of LBM becomes most pronounced at large grid sizes, where communication overhead and global-solve cost dominate in SEM/FVM frameworks.

\begin{figure}[htbp]
			\centering
			\includegraphics[width=0.9\linewidth]{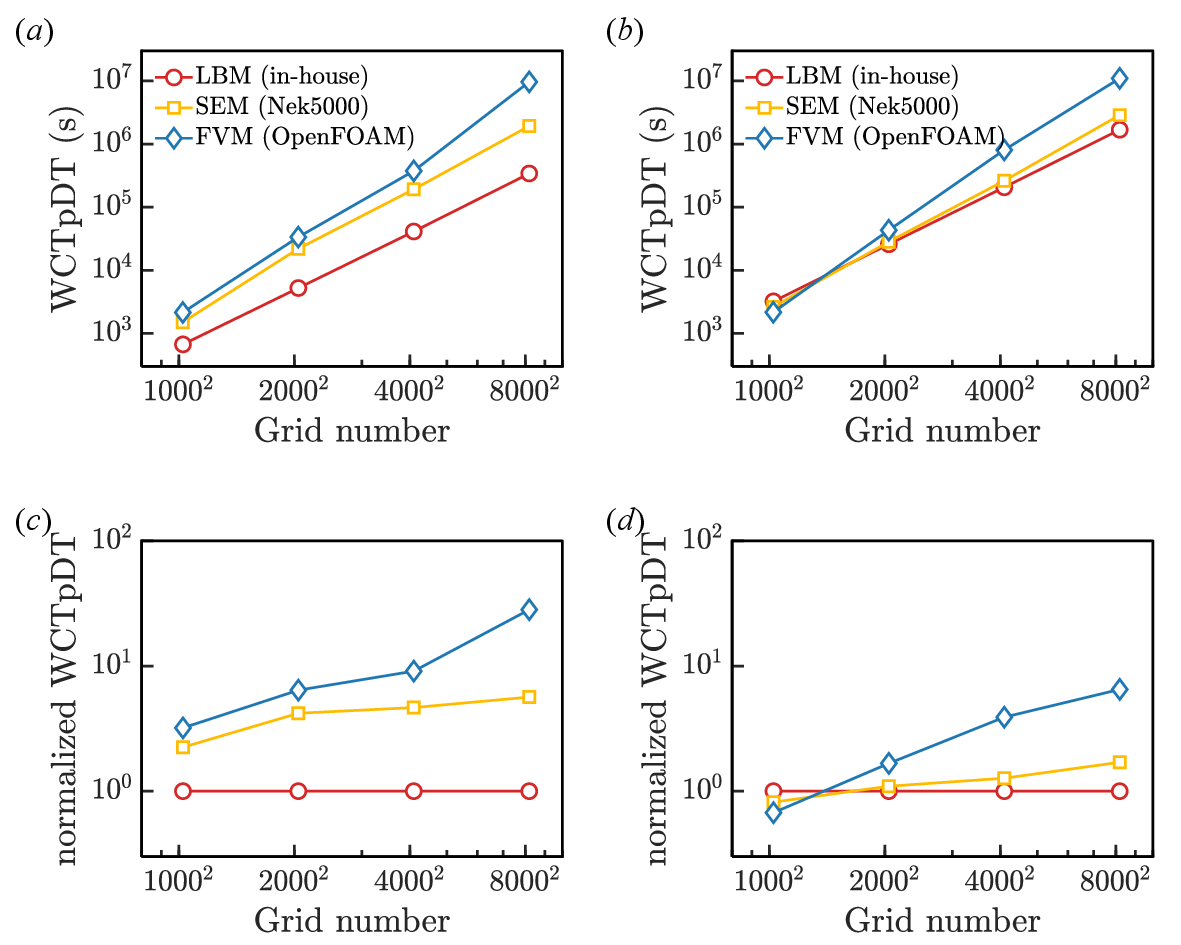}
			\caption{Performance comparison of in-house LBM solver, Nek5000  and OpenFOAM  in terms of wall-clock time per dimensionless time (WCTpDT). 
(\emph{a}, \emph{b}) Absolute WCTpDT as a function of grid number; (\emph{c}, \emph{d}) WCTpDT normalized by the corresponding LBM values,
on (\emph{a}, \emph{c}) uniform and (\emph{b}, \emph{d}) non-uniform meshes. }\label{fig_compare_WCT_2D}
		\end{figure}

For steady problems, different solvers may require varying numbers of time steps to converge, depending on the temporal discretization scheme employed and the inherent numerical damping of each solver. 
In unsteady simulations, however, it is often necessary to quantify the computational cost of advancing the solution over a prescribed dimensionless physical time (e.g., 1, 10, or 100 free-fall time). 
This requirement is particularly relevant in large-scale turbulent convection studies, where long integrations are needed to achieve statistical stationarity and to accumulate converged averages. 
To this end, we introduce the metric WCTpDT (Wall-Clock Time per Dimensionless Time), defined as:
		\begin{equation}
			\mathrm{WCTpDT}=\frac{\text{measured wall-clock time}}{\text{simulated dimensionless time}}\mathrm{~[s]} 
			\label{eq:27}
		\end{equation}
which quantifies the wall-clock time required to simulate one unit of dimensionless time on a given solver and hardware configuration.
A lower WCTpDT indicates greater efficiency, as fewer computational resources are required to advance the solution over one unit of dimensionless time.
Figs. \ref{fig_compare_WCT_2D}(a) and \ref{fig_compare_WCT_2D}(b) show WCTpDT as a function of grid number for uniform and non-uniform meshes, respectively. 
On uniform meshes, the LBM solver consistently achieves the lowest WCTpDT across all tested resolutions. 
For non-uniform meshes, OpenFOAM shows a modest advantage at the coarsest resolution, but as grid size increases, the LBM solver demonstrates superior time-advancement efficiency. 
This advantage becomes even clearer when WCTpDT is normalized by the LBM results, as shown in Figs. \ref{fig_compare_WCT_2D}(c) and \ref{fig_compare_WCT_2D}(d). 
It is worth noting that the Courant-Friedrichs-Lewy (CFL) numbers adopted were 0.5 for Nek5000 and 0.9 for OpenFOAM, allowing these solvers to employ relatively larger time steps than LBM. 
On the uniform mesh, the effective nondimensional timestep is $dt/dx=Ma \cdot c_s \approx 0.0577$ for LBM, compared with $dt/dx \approx 0.8$ for Nek5000 and $dt/dx\approx 1.2$ for OpenFOAM. 
While larger timesteps should, in principle, reduce the cost per simulated physical time, Nek5000 and OpenFOAM exhibit substantially higher per-step overhead.
As a result, their WCTpDT values are roughly an order of magnitude larger than those of LBM. 
This explains why, despite LBM attaining two to three orders of magnitude higher MLUPS values than Nek5000 and OpenFOAM, the relative advantage in WCTpDT is smaller. 
Nevertheless, as the grid resolution increases, the efficiency gap widens and the LBM solver becomes increasingly advantageous.

\begin{figure}[htbp]
			\centering
			\includegraphics[width=0.9\linewidth]{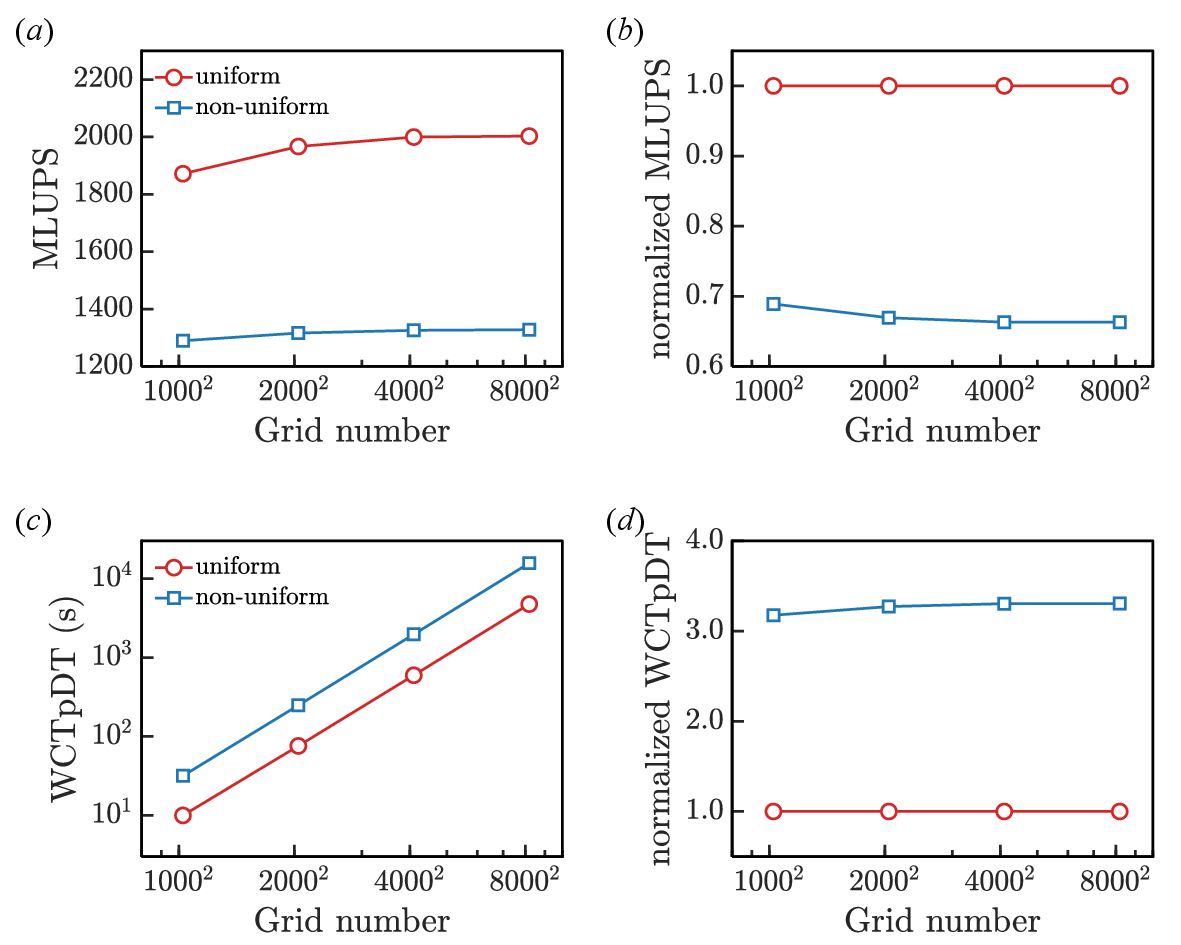}
			\caption{GPU performance of the in-house LBM solver on an NVIDIA A100 for uniform and non-uniform meshes. 
(\emph{a}) Absolute MLUPS as a function of grid number, (\emph{b}) MLUPS normalized by the corresponding uniform-mesh performance. 
(\emph{c}) Absolute WCTpDT as a function of grid number, (\emph{d}) WCTpDT normalized by the corresponding uniform-mesh value.}\label{fig_GPUperformance2D}
		\end{figure}

To further assess the computational efficiency of the in-house LBM solver on modern GPU architectures, additional performance tests were conducted on an NVIDIA A100 GPU. 
Fig. \ref{fig_GPUperformance2D} compares the solver's performance on uniform and non-uniform meshes in terms of MLUPS and WCTpDT across different grid sizes. 
As shown in Fig. \ref{fig_GPUperformance2D}(a), on all tested grids, MLUPS values on uniform meshes consistently exceed those on non-uniform meshes.
This disparity stems primarily from the additional interpolation operations required for off-lattice streaming on non-uniform grids, which introduce irregular memory access patterns and reduce throughput.
The normalized MLUPS (see Fig. \ref{fig_GPUperformance2D}b) shows that non-uniform grids achieve only 60–70\% of the performance of uniform grids, depending on resolution.
The quadratic interpolation step, particularly its memory-bound indirect access pattern, is the main source of this performance degradation.
Uniform meshes permit fully coalesced and predictable memory access, whereas non-uniform interpolation requires fetching data from spatially non-contiguous memory locations, leading to irregular global memory access and reduced throughput.
The collision step, in contrast, is compute-bound and consistently sustains high arithmetic throughput regardless of mesh type.
Consequently, the observed 30–40\% drop in MLUPS for non-uniform grids reflects interpolation overhead rather than any intrinsic inefficiency of the LBM algorithm.
WCTpDT increases nearly linearly with grid size for both mesh types (see Fig. \ref{fig_GPUperformance2D}c), while the relative overhead associated with non-uniform meshes remains stable at about a factor of three across resolutions (see Fig. \ref{fig_GPUperformance2D}d).
These results confirm that, despite modest penalties from interpolation overhead, the in-house LBM solver retains excellent scalability and GPU efficiency on both mesh types.
This enables practical large-scale simulations on contemporary accelerator hardware.
It is also worth noting that our LBM solver on GPUs employs double-precision floating point arithmetic to ensure accuracy.

We also assess the economic benefits of the LBM solver on GPU platforms through a simple cost estimation. 
Consider a simulation at $Ra=10^{8}$ and $Pr=0.71$ using a $8193^2$ non-uniform mesh with a stretching coefficient of $a=2.1$. 
On an NVIDIA A100 GPU, the wall-clock time required to advance the simulation by 1 $t_f$ (i.e. WCTpDT) is $1.5722\times 10^{4} \ \text{s} \approx 4.37 \ \text{h}$.
Assuming a nominal rental cost of 7 RMB per GPU-hour, the corresponding cost per $t_f$ is  $4.37 \ \text{h} \times 7 \ \text{RMB/h}\approx 30.57\ \text{RMB}$. 
By contrast, CPU-based solvers requires significantly longer wall-clock times. 
For Nek5000, the time per $t_f$ is about $2.871852\times 10^6 \ \text{s} \approx  797.74 \ \text{h}$, yielding a cost of $797.74 \ \text{h}\times0.1\ \text{RMB/h}\approx79.77\ \text{RMB}$, where we assume a conservative rate of $0.1\ \text{RMB}$ per CPU-core hour. 
For OpenFOAM, the required time increases to $1.0981646\times 10^7 \ \text{s} \approx 3050.5 \ \text{h}$, yielding $3050.5 \ \text{h}\times0.1\ \text{RMB/h}\approx305.05\ \text{RMB}$. 
It is important to emphasize that both Nek5000 and OpenFOAM typically run on multi-core CPU architectures, where scalability is often limited by communication overhead and memory contention. 
Consequently, the effective performance falls short of ideal expectations, further increasing wall-clock times and overall computational cost. 
These considerations highlight the substantial economic advantage of the GPU-based LBM solver, an advantage that becomes increasingly pronounced in large-scale simulations.

\section{Laminar thermal convection in a 3-D side-heated cavity} \label{sec:SHC-3D}

We next consider thermal convection in the canonical 3-D side-heated cavity \cite{xu2019lattice}. 
The left and right vertical walls are maintained at constant hot and cold temperatures, respectively, while the other four walls are adiabatic. 
All six walls impose no-slip velocity boundary conditions. 
In our previous work \cite{xu2019lattice}, we provided results obtained on a uniform mesh of $257^3$ grid points, which were subsequently verified by 
Sun and Tao \cite{sun2023performance}, Chen et al. \cite{chen2022evolutions}, Ren et al. \cite{Ren2021,Ren2023}, Yigit et al. \cite{Yigit2020}, Vesper et al. \cite{Vesper2022}, and many others \cite{Xu2022,Lu2022,Su2019,Ma2023,Schupbach2025}. 
Here, to validate the ISLBM for simulating coupled fluid flow and heat transfer in 3-D, we present simulation results for $10^5\leq Ra\leq10^7$, with the Prandtl number fixed at 0.71.
Under these conditions, the flow reaches a steady state once the same convergence criteria as in the 2-D case are satisfied (see Eq. (\ref{eq:steadyCriteria})).

\begin{figure}[htbp]
			\centering
			\includegraphics[width=1\linewidth]{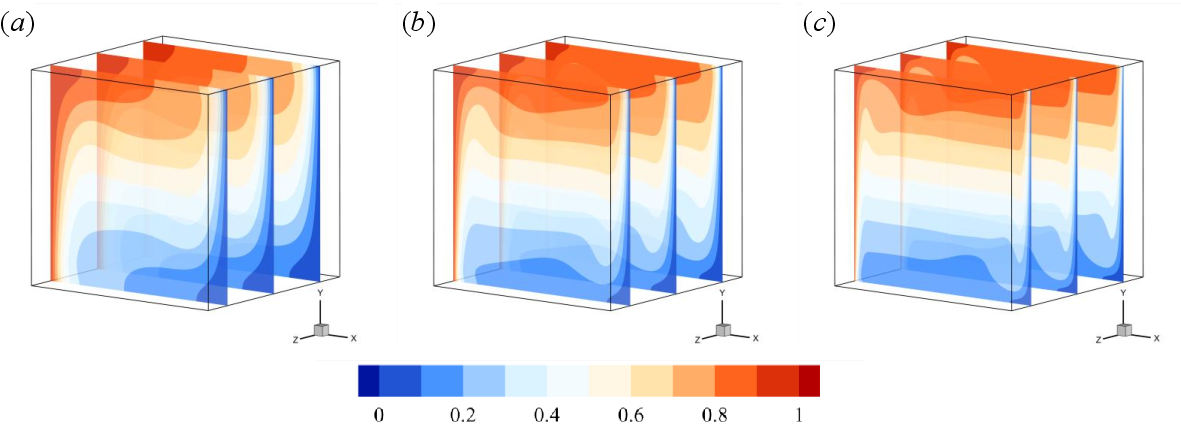}
			\caption{Temperature slices $T^{*}$ at (\emph{a}) $Ra=10^5$, (\emph{b}) $Ra=10^6$, and (\emph{c}) $Ra=10^{7}$.}\label{fig_temperature3D}
		\end{figure}

Fig. \ref{fig_temperature3D} illustrates the temperature fields $T^{*}$, where contour slices in the vertical planes highlight the evolution of thermal structures with increasing buoyancy forcing. 
At $Ra=10^5$ (see Fig. \ref{fig_temperature3D}a), the flow is characterized by relatively smooth and broad thermal gradients with thick boundary layers. 
As $Ra$ increases to $10^6$ (see Fig. \ref{fig_temperature3D}b), convection becomes more vigorous, leading to thinner boundary layers and sharper vertical temperature gradients, indicative of enhanced plume activity. 
At $Ra = 10^7$ (see Fig. \ref{fig_temperature3D}c), convective transport dominates: the temperature field exhibits well-defined plume structures, pronounced vertical mixing, and significantly thinner boundary layers. 
These visualizations demonstrate the progressive transition from conduction-dominated to convection-dominated regimes as $Ra$ increases. 
The overall structure agrees well with earlier 3-D simulations (see Fig. 10 in our previous work \cite{xu2019lattice} and with many other studies \cite{sun2023performance,chen2022evolutions}). 

\begin{figure}[htbp]
			\centering
			\includegraphics[width=0.9\linewidth]{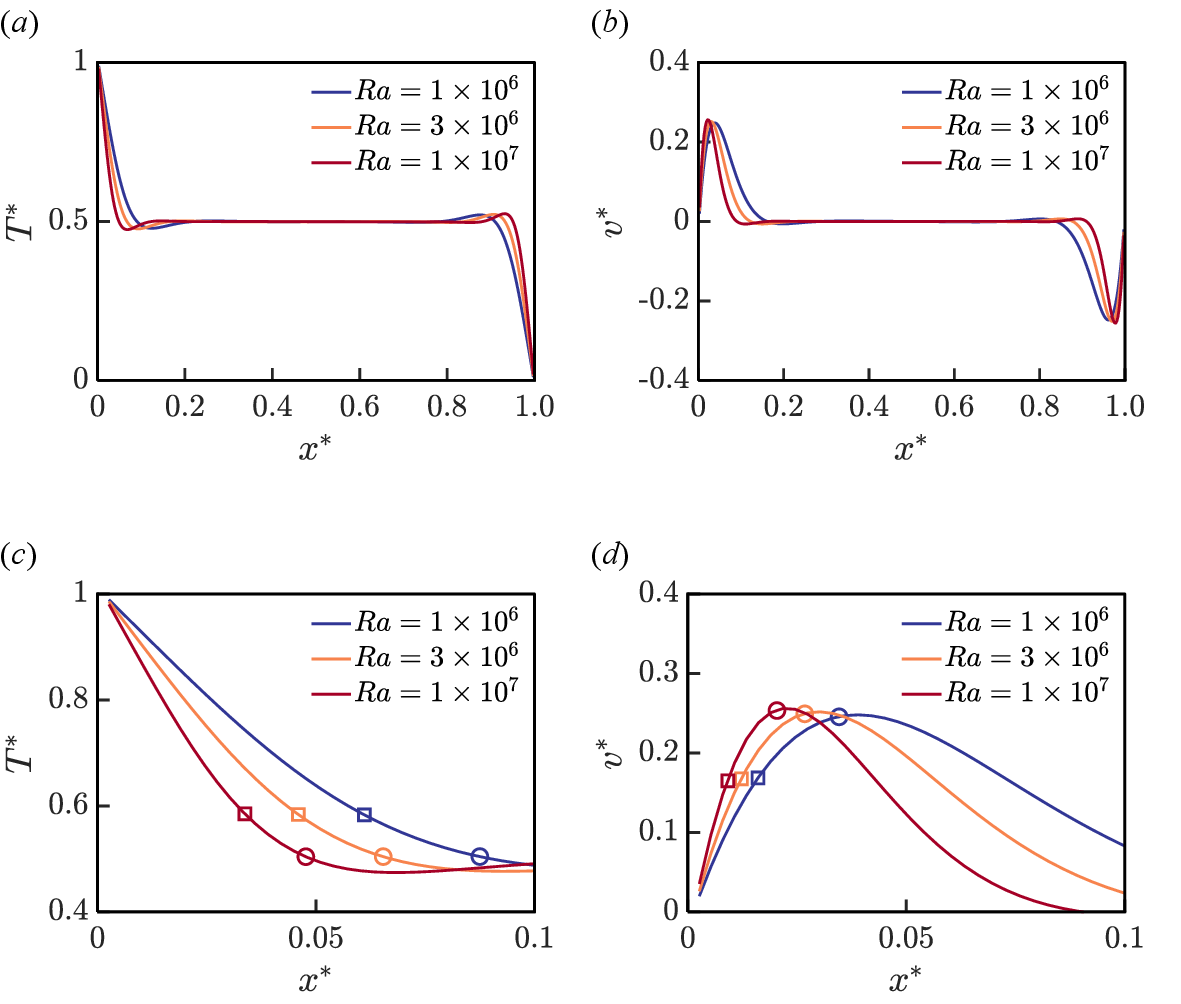}
			\caption{Horizontal distributions of vertically averaged (\emph{a}) temperature $T^{*}$ and (\emph{b}) vertical velocity $v^{*}$.
Panels (\emph{c}, \emph{d}) show enlarged views near the hot wall $(x^{*}=0)$, where open circles denote boundary layer locations determined by the 99\% criterion method, and open squares denote boundary layer locations determined by the slope method.}\label{fig_verticalAvg3D}
		\end{figure}

Figs. \ref{fig_verticalAvg3D}(a) and \ref{fig_verticalAvg3D}(b) show the horizontal distributions of the temperature $T^{*}$ and vertical velocity $v^{*}$, each averaged over the mid-plane band $0.4 \leq y \leq 0.6$ for three Rayleigh numbers. 
Both quantities exhibit clear gradients near the sidewalls, indicating the formation of thermal and momentum boundary layers. 
Enlarged views near the hot wall ($x^{*} = 0$) are provided in Figs. \ref{fig_verticalAvg3D}(c) and \ref{fig_verticalAvg3D}(d), where open circles and squares denote the boundary layer thicknesses determined using the 99\% criterion and slope methods, respectively. 
From these measurements, the scaling relations are obtained as $\delta_T \sim Ra^{-0.26}$ for the thermal boundary layer and $\delta_V \sim Ra^{-0.23}$ for the velocity boundary layer, with consistent results from both methods.	

In Table \ref{tb:3DNuRe}, we present quantitative results for heat transfer efficiency in terms of the Nusselt number, and for global flow strength in terms of the Reynolds number. 
Specifically, the volume-averaged Nusselt number $\langle{Nu}\rangle$ is defined as
		\begin{equation}
			\langle Nu\rangle=\sqrt{RaPr}\langle u^*T^*\rangle_V+1
		\end{equation}
and the average Nusselt numbers along the hot and cold walls, $Nu_{\text{hot}}$ and $Nu_{\text{cold}}$, are given by
		\begin{equation}
			Nu_{\text{hot}}=-\langle\frac{\partial T^*}{\partial x^*}\rangle_{\text{hot}},\quad Nu_{\text{cold}}=-\langle\frac{\partial T^*}{\partial x^*}\rangle_{\text{cold}}
		\end{equation}
The volume-averaged Reynolds number $\langle Re\rangle$ is defined as
		\begin{equation}
			\langle Re\rangle=\sqrt{\frac{Ra}{Pr}}\langle\sqrt{u^{*2}+v^{*2}}\rangle_V
			\label{eq:4}
		\end{equation}
Here, $\langle\cdot\rangle_{V}$ denotes a volume average, while $\langle\cdot\rangle_{\text{hot/cold}}$ denotes an area average along the hot or cold wall. 
For comparison, we also include data from previous studies, such as 
Fusegi et al. \cite{Fusegi1991}, who employed a control-volume-based finite difference method with a strongly implicit scheme; 
Tric et al. \cite{Tric2000}, who used a pseudo-spectral Chebyshev algorithm based on the projection–diffusion method; and
our earlier work \cite{xu2019lattice}, which employed the LBM with Guo’s forcing scheme \cite{guo2002discrete,guo2008analysis,chai2012effect} on a uniform mesh.
		\begin{table}[htbp]
			\centering
			\label{tb:3DNuRe}
			\caption{Benchmark solutions of the Nusselt numbers and Reynolds number. The columns from left to right indicate the following: Rayleigh number $Ra$; flow database; grid number $N_x \times N_y \times N_z$; volume-averaged Nusselt number $\langle Nu \rangle$; average Nusselt number along the hot wall $Nu_{\text{hot}}$; average Nusselt number along the cold wall $Nu_{\text{cold}}$; volume-averaged Reynolds number $\langle Re \rangle$.}
			\resizebox{1\textwidth}{!}{
				\begin{tabular}{ccccccc}
					\toprule
					\makecell[c]{$Ra$} & \makecell[c]{Flow database} & \makecell[c]{$N_x \times N_y \times N_z$} & \makecell[c]{$\langle Nu \rangle$} & \makecell[c]{$Nu_{\text{hot}}$} & \makecell[c]{$Nu_{\text{cold}}$} & \makecell[c]{$\langle Re \rangle$} \\
					\midrule
					
					\multirow[t]{6}{*}{$10^5$} 
					& Present & $129^3$ & 4.3406 & 4.3624 & 4.3624 & 39.4521 \\
					& Present & $181^3$ & 4.3363 & 4.3520 & 4.3520 & 39.3768 \\
					& Present & $257^3$ & 4.3346 & 4.3457 & 4.3457 & 39.3500 \\
					&         & $\infty$ & 4.3334 & 4.3362 & 4.3362 & 39.3352 \\
					&         & $p$ & 2.6 & 1.5 & 1.5 & 3.0 \\[2pt]
					&         & GCI (\%) & 0.03 & 0.3 & 0.3 & 0.05 \\[5pt]
					& Fusegi \cite{Fusegi1991} & $62^3$ & -- & 4.361 & -- & -- \\
					& Tric \cite{Tric2000}   & $81^3$ & --   & --     & 4.3370 & -- \\[1pt]
                    \hline \\[1pt]
	
					\multirow[t]{7}{*}{$10^6$} 
					& Present & $181^3$ & 8.6601 & 8.6931 & 8.6931 & 97.4828 \\
					& Present & $257^3$ & 8.6441 & 8.6669 & 8.6668 & 97.0075 \\
					& Present & $361^3$ & 8.6379 & 8.6540 & 8.6540 & 96.8464 \\
					&         & $\infty$ & 8.6341 & 8.6419 & 8.6419 & 96.7638 \\
					&         & $p$ & 2.8 & 2.1 & 2.1 & 3.1 \\[2pt]
					&         & GCI (\%) & 0.055 & 0.18 & 0.18 & 0.11 \\[5pt]
					& Xu \cite{xu2019lattice} & $257^3$ & -- & 8.64345 & 8.64342 & -- \\
                    & Fusegi \cite{Fusegi1991} & $62^3$ & -- & 8.770 & -- & -- \\
					& Tric \cite{Tric2000}   & $81^3$ & -- & -- & 8.6407 & -- \\[1pt]
                    \hline \\[1pt]
					
					\multirow[t]{7}{*}{$10^7$} 
					& Present & $181^3$ & 16.5341 & 16.6161 & 16.6158 & 237.3925 \\
					& Present & $257^3$ & 16.4179 & 16.4681 & 16.4680 & 230.7625 \\
					& Present & $361^3$ & 16.3670 & 16.4001 & 16.4001 & 228.4148 \\
					&         & $\infty$ & 16.3274 & 16.3422 & 16.3422 & 227.1278 \\
					&         & $p$ & 2.4 & 2.3 & 2.3 & 3.0 \\[2pt]
					&         & GCI (\%) & 0.30 & 0.44 & 0.44 & 0.70 \\[5pt]
					& Xu \cite{xu2019lattice} & $257^3$ & -- & 16.40322 & 16.40285 & -- \\
					& Tric \cite{Tric2000}   & $111^3$ & -- & -- & 16.3427 & -- \\[1pt]
                    					
					\bottomrule
				\end{tabular}%
			}
		\end{table}
		
The results in Table \ref{tb:3DNuRe} demonstrate good agreement with previous studies, with most values converge toward the corresponding asymptotic limits. 
To ensure systematic refinement, the non-uniform mesh was constructed to preserve geometric similarity while reducing the effective mesh spacing. 
Richardson extrapolation was then applied using three successively refined meshes: 
for $Ra=10^5$, the grid numbers are $(N_{\text{coarse}}, N_{\text{medium}}, N_{\text{fine}})=(129,181,361)$; 
for $Ra=10^6$ and $10^7$, they are $(N_{\text{coarse}}, N_{\text{medium}}, N_{\text{fine}})=(181,257,361)$. 
Throughout, we take $\Delta x_i\propto1/N_i$, with a refinement ratio $r=\Delta x_{\text{coarse}}/\Delta x_{\text{medium}}=\Delta x_{\text{medium}}/\Delta x_{\text{fine}}\approx 1.41$. 
Table \ref{tb:3DNuRe} reports the estimated convergence orders and Richardson-extrapolated asymptotic values of the Nusselt and Reynolds numbers. 
The included GCI values confirm that numerical uncertainty is generally small, typically below 1\%. 
Larger uncertainties are observed for local variables, reflecting the greater difficulty of achieving convergence in these quantities. 
Nevertheless, these results validate the robustness of the grid-refinement study.
The ISLBM solver achieves nearly third-order spatial accuracy for volume-averaged Nusselt and Reynolds numbers, owing to the error-canceling effect of domain integration. 
To further examine this behavior, Table \ref{tb:3Duvt} summarizes flow variables evaluated at the cell center. 
Consistent with the 2-D case, these local flow variables converge at approximately second order.

		\begin{table}[htbp]
			\centering
			\caption{Convergence behavior of flow variables at the cell center. The columns from left to right indicate the following: Rayleigh number $Ra$; grid number $N_x \times N_y \times N_z$; absolute value of vorticity magnitude $|\omega_{c}|$; temperature $T_{c}$; horizontal velocity $u_{c}$; vertical velocity $v_{c}$.}
			\label{tb:3Duvt}
			\resizebox{1\textwidth}{!}{
				\begin{tabular}{cccccc}
					\toprule
					\makecell[c]{$Ra$} & \makecell[c] {Grid} & \makecell[c]{$|\omega_{c}|$} & \makecell[c]{$T_{c}$} & \makecell[c]{$u_{c}$} & \makecell[c]{$v_{c}$} \\
					\midrule
					
					\multirow[t]{6}{*}{$10^5$}
					& $129^3$ & 0.2565 & 0.49948 & $-2.1342\times10^{-4}$ & $1.3056\times10^{-4}$ \\
					& $181^3$ & 0.2560 & 0.49949 & $-2.2689\times10^{-4}$ & $1.2423\times10^{-4}$ \\
					& $257^3$ & 0.2557 & 0.49950 & $-2.3377\times10^{-4}$ & $1.2119\times10^{-4}$ \\
					& $\infty$ & 0.2548 & 0.49950 & $-2.4093\times10^{-4}$ & $1.1837\times10^{-4}$ \\
					& $p$ & 0.9 & 2.0 & 2.0 & 2.1 \\
					& GCI (\%) & 0.5 & 0.002 & 4.0 & 3.0 \\[1pt]
                    \hline \\[1pt]
					
					\multirow[t]{6}{*}{$10^6$}
					& $181^3$ & 0.1419 & 0.49942 & $-1.2268\times10^{-4}$ & $1.4722\times10^{-5}$ \\
					& $257^3$ & 0.1383 & 0.49946 & $-1.3361\times10^{-4}$ & $1.6437\times10^{-5}$ \\
					& $361^3$ & 0.1370 & 0.49948 & $-1.3933\times10^{-4}$ & $1.7106\times10^{-5}$ \\
					& $\infty$ & 0.1363 & 0.49949 & $-1.4563\times10^{-4}$ & $1.7532\times10^{-5}$\\ 
					& $p$ & 3.0 & 2.1 & 1.9 & 2.1 \\
					& GCI (\%) & 0.63 & 0.0039 & 5.7 & 3.1 \\[1pt] 
                    \hline \\[1pt]
					
					\multirow[t]{6}{*}{$10^7$}
					& $181^3$ & 0.1296 & 0.49898 & $-7.1568\times10^{-5}$ & $6.1396\times10^{-6}$ \\
					& $257^3$ & 0.1149 & 0.49917 & $-7.3572\times10^{-5}$ & $7.4064\times10^{-6}$ \\
					& $361^3$ & 0.1094 & 0.49926 & $-7.8387\times10^{-5}$ & $7.7847\times10^{-6}$ \\
					& $\infty$ & 0.1060 & 0.49934 & -- & $7.9458\times10^{-6}$ \\
					& $p$ & 2.8 & 2.3 & -- & 3.5 \\
					& GCI (\%) & 3.9 & 0.019 & -- & 2.6 \\
					
					\bottomrule
				\end{tabular}%
			}
		\end{table}
		
\begin{figure}[htbp]
			\centering
			\includegraphics[width=1\linewidth]{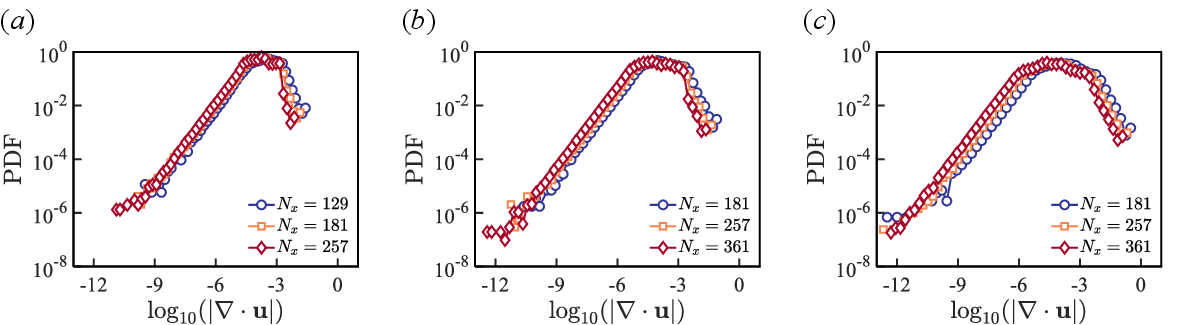}
			\caption{PDFs of $\log_{10}(|\nabla\cdot\mathbf{u}|)$ obtained over the entire domain, for (\emph{a}) $Ra=10^5$, (\emph{b}) $Ra=10^6$, and (\emph{c}) $Ra=10^7$.}\label{fig_divU3D}
		\end{figure}

As in the 2-D case, compressibility effects are assessed by examining the divergence field in three dimensions. 
Fig. \ref{fig_divU3D} shows the PDFs of $\log_{10}(|\nabla\cdot\mathbf{u}|)$ at $Ra=10^5$, $10^6$, and $10^7$, evaluated at three different grid resolutions. 
Across all Rayleigh numbers, the PDFs peak between $|\nabla\cdot\mathbf{u}|\in [ 10^{-6},10^{-4} ]$ with long tails extending toward $10^{-11}$, indicating that the flow remains nearly solenoidal. 
With increasing resolution, the distributions collapse onto each other, demonstrating numerical convergence of the divergence statistics. 
At higher Rayleigh numbers, the PDFs broaden slightly and the mean divergence increases modestly, reflecting stronger thermal forcing and sharper gradients that locally amplify compressibility effects. 
Nevertheless, these deviations remain small in magnitude.
Taken together with the 2-D results, these findings confirm that the weakly compressible LBM preserves the incompressible character of thermal convection in both two and three dimensions, even at high Rayleigh numbers.

\begin{figure}[htbp]
			\centering
			\includegraphics[width=0.9\linewidth]{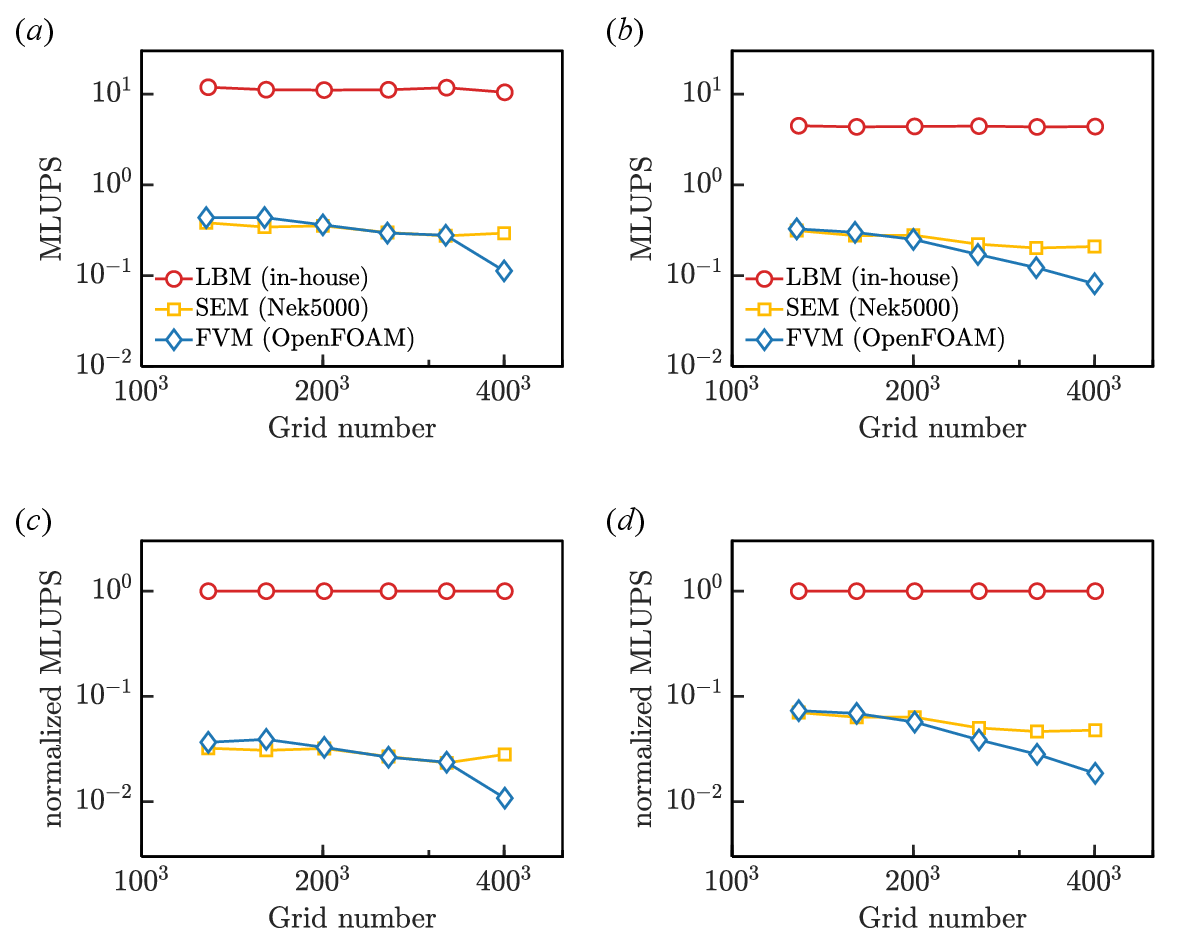}
			\caption{Performance comparison of in-house LBM solver, Nek5000 and OpenFOAM in terms of MLUPS. 
(\emph{a}, \emph{b}) Absolute MLUPS as a function of grid number, (\emph{c}, \emph{d}) MLUPS normalized by the corresponding LBM solver, on (\emph{a}, \emph{c}) uniform and (\emph{b}, \emph{d}) non-uniform meshes.}\label{fig_compare_MLUPS_3D}
		\end{figure}

To extend the performance assessment to three dimensions, we adopt MLUPS as the performance metric. 
Fig. \ref{fig_compare_MLUPS_3D} shows results for 3-D simulations at $Ra=10^7$ and $Pr=0.71$ on both uniform and non-uniform meshes executed on AMD EPYC 9135 CPUs.
For non-uniform meshes, the stretching coefficient $a=2.1$. 
As seen in  Figs. \ref{fig_compare_MLUPS_3D}(a) and \ref{fig_compare_MLUPS_3D}(b), the absolute MLUPS values reveal that the in-house LBM solver consistently sustains throughput at $O(10)$ MLUPS across all tested grid sizes, while Nek5000 and OpenFOAM are two to three orders of magnitude slower and exhibit decreasing performance as the grid size increases. 
On the largest grids, the throughput of Nek5000 and OpenFOAM drops to near $O(10^{-1})$ MLUPS, whereas the in-house LBM solver maintains nearly constant performance, demonstrating resilience to increasing problem size. 
The normalized MLUPS results in Figs. \ref{fig_compare_MLUPS_3D}(c) and \ref{fig_compare_MLUPS_3D}(d) further highlight the relative efficiency of the three solvers: 
Nek5000 and OpenFOAM achieve at most a few percent of the LBM performance, with the performance gap widening at higher resolutions, particularly on non-uniform meshes. 
By contrast, the relative advantage of the LBM solver remains robust or even improves as the grid size increases.
 
\begin{figure}[htbp]
			\centering
			\includegraphics[width=0.9\linewidth]{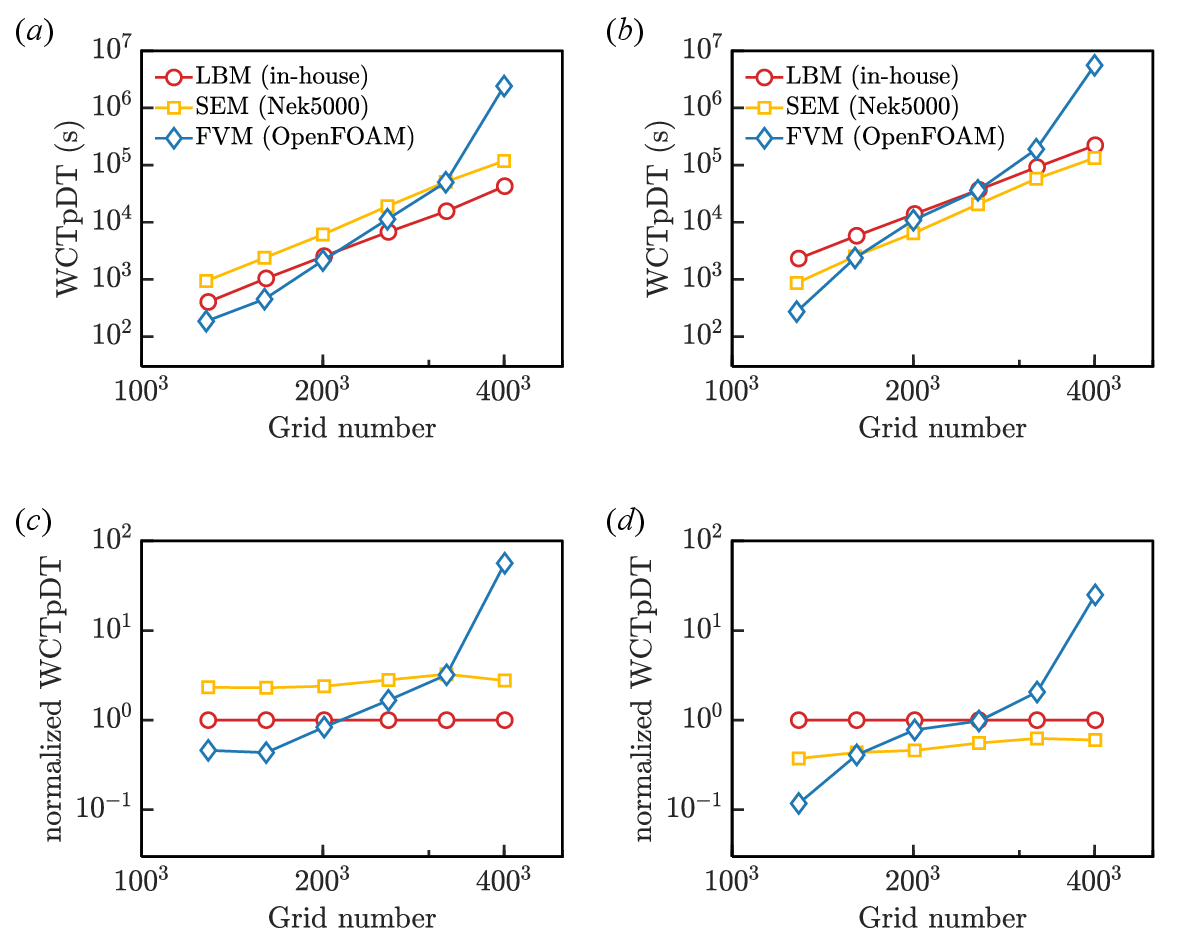}
			\caption{Performance comparison of in-house LBM solver,  Nek5000 and OpenFOAM solvers in terms of WCTpDT. 
(\emph{a}, \emph{b}) Absolute WCTpDT as a function of grid number, (\emph{c}, \emph{d}) the WCTpDT normalized by the corresponding LBM solver, on (\emph{a}, \emph{c}) uniform and (\emph{b}, \emph{d}) non-uniform meshes.}\label{fig_compare_WCT_3D}
		\end{figure}

Figs. \ref{fig_compare_WCT_3D}(a) and \ref{fig_compare_WCT_3D}(b) further show the absolute WCTpDT values as functions of grid number on uniform and non-uniform meshes, respectively. 
On uniform meshes, the in-house LBM solver consistently achieves shorter runtimes than Nek5000, while OpenFOAM is slightly more efficient at the coarsest grid but quickly loses this advantage as the resolution increases, eventually becoming the most expensive solver. 
On non-uniform meshes, OpenFOAM again performs best at the smallest grid, but its cost grows steeply with resolution, surpassing both LBM and Nek5000 by more than an order of magnitude at the largest grids. 
Nek5000, by contrast, scales more smoothly and maintains a modest runtime advantage over LBM across all non-uniform grid sizes. 
The normalized results in Figs. \ref{fig_compare_WCT_3D}(c) and \ref{fig_compare_WCT_3D}(d) further highlight these trends. 
On uniform meshes, Nek5000 requires approximately 3–5 times the runtime of LBM per unit dimensionless time, while OpenFOAM starts faster at coarse grids but becomes significantly more expensive at higher resolutions. 
On non-uniform meshes, Nek5000 consistently outperforms LBM, requiring only about 50–80\% of its runtime, whereas OpenFOAM undergoes a sharp transition from being the fastest solver at small grids to the least efficient at the large grids, exceeding LBM's cost by a factor of ~5–6. 
Although Nek5000 and OpenFOAM employ larger CFL numbers (0.5 and 0.9, respectively) than the LBM solver, which should in principle reduce their cost per simulated physical time,  their substantially higher per-step computational overhead offsets this benefit.
Consequently, WCTpDT values for both remain larger, particularly at high resolutions. 
As the grid size increases, the efficiency gap widens further, establishing the LBM solver as a favorable choice for advancing large-scale 3-D simulations.

\begin{figure}[htbp]
			\centering
			\includegraphics[width=0.9\linewidth]{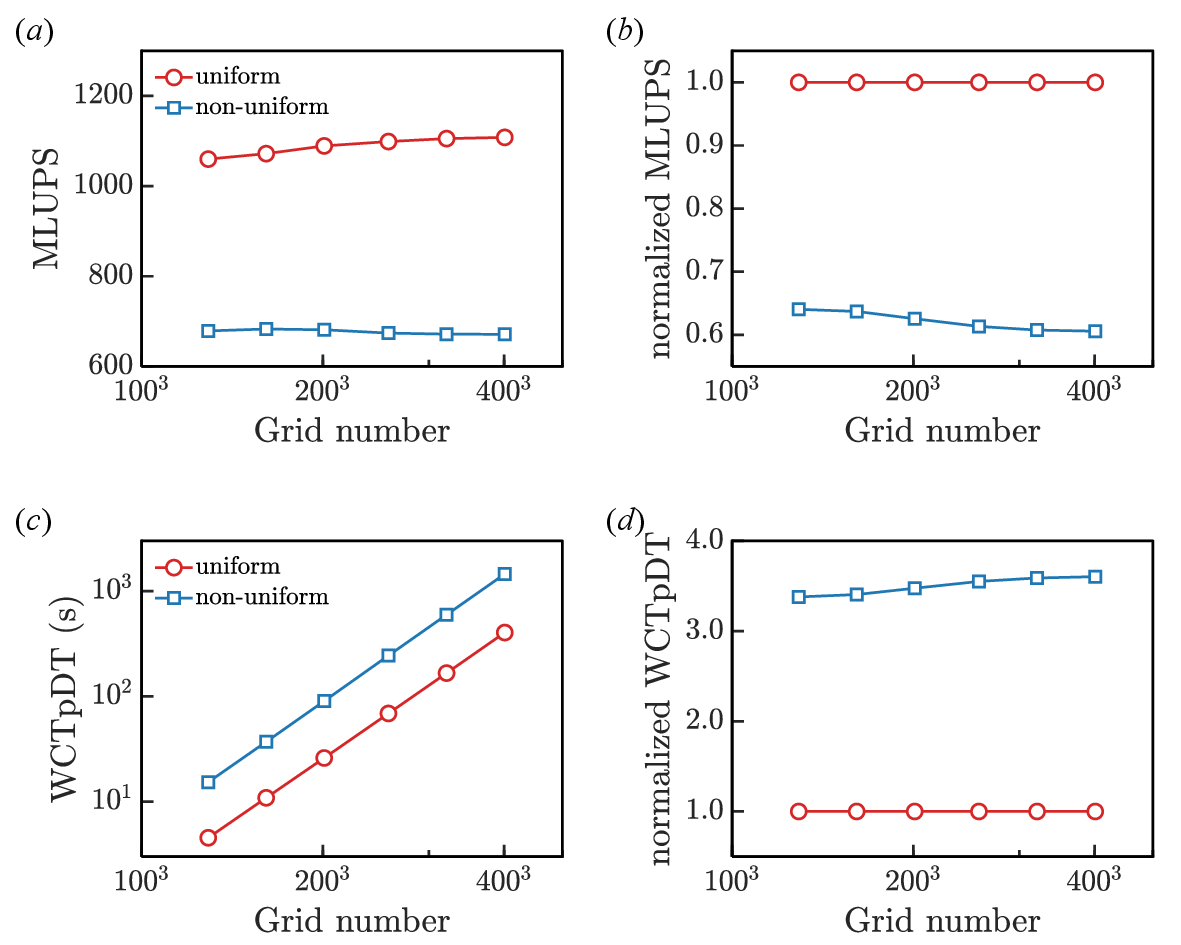}
			\caption{GPU performance of the in-house LBM solver on an NVIDIA A100 for uniform and non-uniform meshes: 
(\emph{a}) Absolute MLUPS as a function of grid number, (\emph{b}) MLUPS normalized by the uniform-mesh performance at each grid; 
(\emph{c}) WCTpDT as a function of grid number; (\emph{d}) WCTpDT normalized by the corresponding value for uniform meshes.}\label{fig_GPUperformance}
		\end{figure}

To further assess the computational efficiency of the in-house 3-D LBM solver on modern GPU architecture, we performed additional tests on an NVIDIA A100 GPU. 
Fig. \ref{fig_GPUperformance} compares the solver's performance for uniform and non-uniform meshes. 
As shown in Fig. \ref{fig_GPUperformance}(a), the MLUPS values on uniform meshes consistently exceed those on non-uniform meshes across all tested grid sizes. 
The normalized MLUPS in Fig. \ref{fig_GPUperformance}(b) indicates that non-uniform grids achieve only about 60–65\% of the performance of uniform grids, with a slight downward trend as resolution increases. 
Fig. \ref{fig_GPUperformance}(c) shows that the WCTpDT grows nearly linearly with grid size for both mesh types, while Fig. \ref{fig_GPUperformance}(d) confirms that the relative overhead of non-uniform meshes remains close to a factor of three across resolutions. 
Compared with the 2-D case, the relative throughput loss is similar, but the absolute runtime penalty is more pronounced in 3-D, reflecting the higher computational and memory demands of large-scale simulations. 
Overall, the solver retains excellent scalability and GPU efficiency in three dimensions, ensuring its applicability to practical large-scale problems even when non-uniform discretizations are employed.

For 3-D simulations, the cost estimation is as follows.
Consider a simulation at $Ra=10^{7}$ and $Pr=0.71$ using a $400^3$ non-uniform mesh with a stretching coefficient of $a=2.1$. 
On an NVIDIA A100 GPU, the wall-clock time required to advance the simulation by 1 $t_f$ (i.e. WCTpDT) is $1.456\times 10^{3} \ \text{s} \approx 0.40 \ \text{h}$. 
Thus, the corresponding cost per $t_f$ is  $0.40 \  \text{h}\times7\ \text{RMB/h}\approx 2.83\ \text{RMB}$. 
For Nek5000, the time per $t_f$ is about $1.33484\times 10^5 \ \text{s} \approx  37.08 \ \text{h}$, yielding a cost of $37.08 \text{h}\times0.1\ \text{RMB/h}\approx3.71\ \text{RMB}$. 
For OpenFOAM, the time per $t_f$ is about $5.576363\times 10^6 \ \text{s} \approx 1549.0 \ \text{h}$, yielding $1549.0 \ \text{h}\times0.1\ \text{RMB/h}\approx154.90\ \text{RMB}$. 
When Nek5000 and OpenFOAM run on multi-core CPU architectures, scalability is limited by communication overhead and memory contention, and their effective performance falls short of ideal expectations, resulting in increased wall-clock times and overall computational cost. 
We expect greater economic advantage of the GPU-based LBM solver in large-scale simulations.

\section{Conclusion} \label{sec:conclusion}

In this study, we systematically evaluated an interpolation-supplemented lattice Boltzmann method (ISLBM) for simulating buoyancy-driven thermal convection on non-uniform meshes.
The ISLBM enables local mesh refinement near solid boundaries to resolve steep thermal and velocity gradients, while retaining the computational efficiency of the standard LBM in the bulk flow. 
By incorporating quadratic interpolation during the streaming step, the ISLBM achieves nearly third-order accuracy for global quantities and about second-order for local fields.
The method was validated through benchmark simulations of laminar thermal convection in a side-heated cavity at Rayleigh numbers $10^6\leq Ra\leq 10^8$ in 2-D and $10^5\leq Ra\leq 10^7$ in 3-D, demonstrating excellent agreement with high-fidelity reference data. 
The results demonstrate second-order spatial accuracy for local temperature and velocity fields, and nearly third-order accuracy for vorticity and volume-integrated quantities. 
Furthermore, a careful assessment of compressibility effects confirmed that the solver remains within a nearly incompressible regime, with only minimal numerical divergence even at elevated Rayleigh numbers.

We also assessed the computational performance of the in-house LBM solver by benchmarking it against two widely used open-source solvers: Nek5000 based on the spectral element method, and OpenFOAM based on the finite volume method. 
Using performance metrics such as MLUPS and WCTpDT, our results consistently demonstrate that the ISLBM outperforms these solvers in large-scale simulations. 
When deployed on GPU architecture, the ISLBM maintains high computational performance. 
In terms of MLUPS, the throughput on non-uniform meshes reaches approximately 60–70\% of that achieved on uniform meshes; 
in terms of WCTpDT, however, the computational cost on non-uniform meshes is about three times higher than that on uniform meshes, mainly due to the overhead of quadratic interpolation and irregular memory access patterns.
Despite this penalty, the solver exhibits excellent scalability and GPU efficiency, enabling practical large-scale simulations on contemporary accelerator hardware.

While the ISLBM was originally proposed by He et al. \cite{he1996some,he1997some,he1997error}, to the best of our knowledge no prior work has conducted a systematic and quantitative evaluation of its performance for thermally driven flows at high Rayleigh numbers on non-uniform meshes. 
This study is, to our knowledge, the first to: 
(i) implement quadratic interpolation-based streaming to simulate thermal convection on non-uniform meshes at high Rayleigh numbers; 
(ii) quantitatively evaluate accuracy, convergence order, divergence control, and computational efficiency; and
(iii) benchmark the ISLBM against high-fidelity solvers (Nek5000 and OpenFOAM) using rigorous MLUPS and WCTpDT metrics, including GPU-based performance tests. 
Our results demonstrate that the interpolation-supplemented LBM provides a robust and efficient framework for simulating thermally driven flows on non-uniform meshes. 
Its demonstrated accuracy, stability, computational efficiency, and scalability make it a promising candidate for future extensions to fully turbulent convection at extreme Rayleigh numbers.

\section*{Acknowledgements}
This work was supported by the National Natural Science Foundation of China (NSFC) through grants nos. 12272311, 12388101; the Young Elite Scientists Sponsorship Program by CAST (2023QNRC001). 
The authors acknowledge the Beijing Beilong Super Cloud Computing Co., Ltd for providing HPC resources that have contributed to the research results reported within this paper (URL: http://www.blsc.cn/).
	
		\bibliographystyle{elsarticle-num}
		

	\end{document}